\documentclass[traditabstract]{aa}
\usepackage{graphicx}
\usepackage{txfonts}
\usepackage{epsfig}
\usepackage{natbib}
\usepackage{color}
\usepackage{rotating}
\usepackage{ulem}

\newcommand{\oversim}[2]{\protect{\mbox{\lower0.5ex\vbox{%
  \baselineskip=0pt\lineskip=0.2ex
  \ialign{$\mathsurround=0pt #1\hfil##\hfil$\crcr#2\crcr\sim\crcr}}}}}
\newcommand{\simless} {\mbox{$\,\mathrel{\mathpalette\oversim<}\,$}} % <~ sign
\usepackage{lineno}
\begin{document}

\title{Local-Group tests of dark-matter concordance cosmology}
\subtitle{Towards a new paradigm for structure formation}
\author{P.~Kroupa\inst{1}, B.~Famaey\inst{1,2}\thanks{Alexander von
    Humboldt Fellow}, K.~S.~de~Boer\inst{1}, J.~Dabringhausen\inst{1},
  M.S.~Pawlowski\inst{1}, C.M.~Boily\inst{2}, H.~Jerjen\inst{3},
  D.~Forbes\inst{4}, G.~Hensler\inst{5}, 
  and M.~Metz\inst{1}\thanks{now at the Deutsches Zentrum f\"ur Luft- und
    Raumfahrt, K\"onigswinterer Str. 522-524, 53227 Bonn, Germany}}
\institute{ Argelander Institute for Astronomy, University of Bonn,
  Auf dem H\"ugel 71,D-53121 Bonn,
  Germany\\ 
\email{pavel|deboer|joedab|mpawlow@astro.uni-bonn.de, manuel.metz@dlr.de}
\and Observatoire Astronomique, Universit\'e de Strasbourg, CNRS UMR
7550, F-67000
Strasbourg, France\\ \email{benoit.famaey|christian.boily@astro.unistra.fr}
\and Research School of Astronomy and Astrophysics, ANU, Mt. Stromlo
Observatory, Weston ACT2611, Australia\\ \email{jerjen@mso.anu.edu.au}
\and Centre for Astrophysics \& Supercomputing, Swinburne University,
Hawthorn VIC 3122, Australia\\ \email{dforbes@swin.edu.au}
\and Institute of Astronomy, University of Vienna,
T\"urkenschanzstr.~17, A-1180 Vienna,
Austria\\ \email{hensler@astro.univie.ac.at}
%
%
%\and Deutsches Zentrum f\"ur Luft- und Raumfahrt, Knigswinterer
%Str. 522-524, 53227 Bonn, Germany\\ \email{manuel.metz@dlr.de}
%
}
\date{{\bf accepted for publication}} 

\abstract{Predictions of the concordance cosmological model (CCM) of
  the structures in the environment of large spiral galaxies are
  compared with observed properties of Local Group galaxies.  Five new
  most probably irreconcilable problems are uncovered:
  1) A great variety
  of published CCM models consistently predict some form of
  relation between dark-matter-mass and luminosity for the Milky Way (MW)
  satellite galaxies, but none is observed.
  2) The mass function of
  luminous sub-halos predicted by the CCM contains too few satellites
  with dark matter (DM) mass $\approx 10^7\,M_\odot$ within their
  innermost 300~pc than in the case of the MW satellites.
  3) The Local Group
  galaxies and data from extragalactic surveys indicate there is a 
  correlation between bulge-mass and the number of
  luminous satellites that is not predicted by the CCM.
  4) The~13 new ultra-faint MW satellites define a
  disc-of-satellites (DoS) that is virtually identical to the DoS
  previously found for the~11 classical MW satellites, implying that most
  of the~24~MW satellites are correlated in phase-space.
  5) The occurrence of two MW-type DM halo masses hosting MW-like galaxies
  is unlikely in the CCM. However, 
  the properties of the Local Group galaxies 
provide information leading to a solution of the above problems.
  The DoS and
  bulge--satellite correlation suggest that dissipational events
  forming bulges are related to the processes forming phase-space
  correlated satellite populations.  These events are well known to
  occur since in galaxy encounters energy and angular momentum are
  expelled in the form of tidal tails, which can fragment to form
  populations of tidal-dwarf galaxies (TDGs) and associated star
  clusters.  If Local Group satellite galaxies are to be interpreted
  as TDGs then the substructure predictions of the CCM are internally in
  conflict.  All findings thus suggest that the CCM does not account
  for the Local Group observations and that therefore existing as well
  as new viable alternatives have to be further explored. These are
  discussed and natural solutions for the above problems emerge.

} 

\titlerunning{Local Group tests of cosmology} 

\authorrunning{Kroupa et al.}  

\maketitle

\section{Introduction}
\label{sec:introd}

Our understanding of the cosmological world relies on two fundamental
assumptions: 1) The validity of General Relativity, and 2)
conservation of matter since the Big Bang. Both assumptions yield the
concordance cosmological model (CCM), according to which an
  initial inflationary period is followed by (exotic, i.e.,
non-baryonic) dark-matter (DM) structures forming and then accreting
baryonic matter, which fuels star formation in the emerging galaxies,
and according to which dark energy (represented by a cosmological
constant $\Lambda$) drives the acceleration of the Universe at a later
epoch. One important way to test assumption~(1) is to compare the
phase-space properties of the nearest galaxies with the expectations
of the CCM. These tests are the focus of the present contribution.

The possibility of the existence of DM was considered more than
85 years ago \citep{Einstein21,Oort32,Zwi33}, and has been under heavy
theoretical and experimental scrutiny \citep{Bertone05} since the
discovery of non-Keplerian galactic rotation curves by \cite{RF70} and
their verification and full establishment by \cite{Bosma81}. The
existence of DM is popularly assumed because it complies with the
General Theory of Relativity, and therefore Newtonian dynamics, in the
weak-field limit. Newtonian dynamics is the simplest form of
gravitational dynamics given that the equations of motion are linear
in the potential, and is thus readily accessible to numerical
simulations of cosmic evolution, upon which the concordance
scenario of structure formation is based \citep{BFPR84}.

The concordance bottom-up scenario of structure formation involving the
repeated accretion of clumps of cold dark matter (CDM) is assumed to
operate throughout the Universe on all scales.  CDM particles with
masses of order of about~$100\,$GeV are the preferred candidates to
account for constraints placed on the matter density, $\Omega_M$, of
thermal relics with realistic cross-sections (see, e.g., eq.~28 of
\citealt{Bertone05}). For lighter particle candidates, the damping
scale becomes too large: for instance, a  hot DM (HDM) particle
candidate ($m_{\rm HDM} \approx$ few eV) would have a free-streaming
length of $\approx 100\,$Mpc leading to too little power at the
small-scale end of the matter power spectrum. The existence of
galaxies at redshift $z \approx 6$ implies that the
coherence scale should have been smaller than $100\,$kpc or so, meaning 
that warm DM (WDM) particles with mass $m_{\rm WDM}\approx 1 - 10\,$keV are
close to being ruled out \citep{Peacock03}.

CDM is a concept that, together with the cosmological constant
($\Lambda$), has been motivated primarily by large-scale observations
of, e.g., the cosmic microwave background (CMB) radiation (WMAP,
\citealt{Spergel07, Komatsu09}), the accelerating universe
(\citealt{Riess98, Perlmutter99}), or the power spectrum of density
perturbations from the SDSS \citep{Tegmark04} and the 2dF galaxy
redshift survey \citep{Cole05}, all of which serve as empirical
benchmarks for calibrating and constraining theoretical scenarios and
cosmological models.  This concordance $\Lambda$CDM model is
consistent with observations on the~Gpc to~Mpc scales
\citep{Reyesetal2010}, but it implies that the Universe evolves towards an
infinite energy content\footnote{One may refer to this issue as the
  ``cosmological energy catastrophy'' in allusion to the black body UV
  catastrophy, which led Max Planck to heuristically introduce an
  auxiliary ($=$ {\sl \uline{H}ilfsgr\"o{\ss}e} in German) number $h$,
  to reproduce the black body spectrum.}  due to the creation of
vacuum energy from dark-energy-driven accelerated expansion
(e.g. \citealt{Peacock99})\footnote{Energy conservation is a
  problematical issue in General Relativity (GR). The
  stress-momentum-energy tensor is a pseudo tensor and so is not
  invariant under a transformation to a different coordinate system
  and back. This may perhaps be considered to indicate that GR may not
  be complete.}. Less problematically perhaps, but nevertheless
noteworthy, the DM particle cannot be contained in the Standard Model
of particle physics without implying a significant revision of
particle physics (e.g. \citealt{Peacock99}).  Strong evidence for the
existence of DM has been suggested by the observations of the
interacting galaxy-cluster pair 1E0657-56 (the ``Bullet cluster'',
\citealt{Clowe06}). The velocity of the sub-cluster relative to
the large cluster has since been calculated to be about 3000~km\,s$^{-1}$ 
so that the observed morphology can 
arise \citep{MB08}. But according to
\cite{AM08} and \cite{LeeKomatsu10}, such high velocities between a
sub-cluster and a main galaxy cluster are virtually excluded in the
CCM. Near the centre of lens-galaxies, the observed delay times
between the multiple images of strongly lensed background sources
cannot be understood if the galaxy has a standard (NFW or isothermal)
DM content and if, at the same time, the Hubble constant has a classical
value of 70 km\,s$^{-1}$\,Mpc$^{-1}$: the solution is either to decrease
the Hubble constant (in disagreement with other observations), or to
consider the known baryonic matter (with constant mass-to-light ratio)
as the one and only source of the lensing \citep{KS04}. On Local
Volume scales (within about 8~Mpc), it has been pointed out that the
Local Void contains far fewer dwarf galaxies than expected if the CCM
were true. At the same time, there are too many large galaxies in the
less crowded parts such that the arrangement of massive galaxies in
the Local Volume is less than 1~per cent likely in the CCM
\citep{PN10}.

This discussion highlights that there are important unsolved issues in
the CCM. This clearly means that substantial effort is
required to understand the problems, to perhaps distill additional 
clues from the data that can provide solutions, and to improve the
theory.

Galaxy formation and evolution is a process that happens on scales
much smaller than 1~Mpc.  Ironically, a major limitation of our
ability to develop a physically consistent model of how galaxies
evolved out of the dark comes from incomplete knowledge of the Local
Group, in particular from the lack of understanding of the structure
and distribution of dwarf satellite galaxies. But, over the past few
years, a steady flow of new results from nearby galaxies including the
Milky Way (MW) and the improving numerical resolution
of computational studies of galaxy formation have allowed ever
more rigorous tests of the CCM.

According to the DM hypothesis, galaxies must have assembled by means of 
accretion and numerous mergers of smaller DM halos. Therefore,
galaxies such as the MW should be swarmed by hundreds to thousands of
these halos \citep{Mooreetal99,Diemand08}, whereby the number of
sub-halos is smaller in WDM than in CDM models
\citep{Knebeetal08}. Furthermore, the triaxial nature of the flow of
matter at
formation would make it impossible to destroy halo substructure by
violent relaxation \citep{Boilyetal04}.  These sub-halos should be
distributed approximately isotropically about their host, and have a
mass function such that the number of sub-halos in the mass interval
$M_{\rm vir}, M_{\rm vir}+dM_{\rm vir}$ is approximately $dN \propto
M_{\rm vir}^{-1.9}\,dM_{\rm vir}$ \citep{Gaoetal04}.

In contrast to this expectation, only a few dozen shining satellites
have been found around both the MW and Andromeda (M31), while the
next largest disc galaxy in the Local Group, M33, has no known satellites. 
The MW hosts the 11 ``classical'' (brightest) satellites, while 13
additional ``new'' and mostly ultra-faint satellite galaxies have been
discovered in the past 15 years primarily by the Sloan Digital
Sky Survey (SDSS)\footnote{For convenience, the
  11~brightest satellite galaxies are here referred to as the
  ``classical'' satellites because these were known before the SDSS
  era. These include the LMC and the SMC with the others being dwarf
  spheroidals.  The other, more recently discovered satellites are
  fainter than the faintest ``classical'' satellites (UMi and Draco),
  and these are called the ``new'' or the ``ultra-faint'' satellites
  or dwarfs (see Table~\ref{tab:satellites}).}.  While the MW
satellites are distributed highly anisotropically (e.g.
\citealt{Klimentowskietal09}), observations of the internal kinematics
(velocity dispersion) of the satellites suggest they are the most DM
dominated galaxies known (e.g. fig.~15 in \citealt{SimonGeha07}). That
is, the velocity dispersions of their stars seem to be defined by an
unseen mass component: the stars are moving faster than can be
accounted for by their luminous matter. The known satellites may
therefore be the luminous ``tip of the iceberg'' of the vast number of
dark sub-halos orbiting major galaxies such as the MW.

Much theoretical effort has been invested in solving the problem that
the number of luminous satellites is so much smaller than the number
of DM-halos predicted by the currently favoured concordance
$\Lambda$CDM hypothesis: stellar feedback and heating processes limit
baryonic growth, re-ionisation stops low-mass DM halos from accreting
sufficient gas to form stars, and tidal forces from the host halo
limit growth of the DM sub-halos and lead to truncation of DM
sub-halos \citep{DekelSilk86, DekelWoo03, MKM, Koposovetal09, OF09,
  Kirby09, Shaya2009, Busha09, Maccioetal09}.  This impressive and
important theoretical effort has led to a detailed quantification of
the DM-mass--luminosity relation of MW satellite galaxies.  Moreover,
the discovery of new (ultra-faint) dSph satellites around the MW
suggests the validity of the ``tip of the iceberg'' notion.  These
lines of reasoning have generally led to the understanding that within
the $\Lambda$CDM cosmology, no serious small-scale issues are apparent
(e.g. \citealt{Tollerudetal2008,Primack09}).

In this contribution we test whether the CCM can be viewed as a
correct description of the Universe by studying generic properties of
the Local Group\footnote{Useful reviews of the Local Group are
  provided by \cite{Mateo98} and \cite{vandenBergh99}.}, which is a
typical environment for galaxies -- the Local Group properties {\sl
  must} conform to the CCM if it is to be valid universally.  To test
this hypothesis, we critically examine state-of-the art models
calculated within the CDM and WDM framework by a number of independent
research groups developed to explain the properties of the faint
satellite galaxies, by comparing them with the following observations:
the mass-luminosity relation for dSph satellites of the Milky Way
(Sect.~\ref{sec:ML}); the mass-distribution of luminous-satellite
halo-masses (Sect.~\ref{sec:mfn}); and the observed relation between
the bulge mass of the host galaxy and the number of satellites
(Sect.~\ref{sec:origin}). The question of whether the
Disc-of-Satellites (DoS) exists, and if in fact the latest MW
satellite discoveries follow the DoS, or whether the existence of the
DoS is challenged by them, is addressed in Sect.~\ref{sec:DoS}.  In
Sect.~\ref{sec:DoS}, the observed invariance of late-type baryonic
galaxies is also discussed in the context of the Local Group.  In
these sections it emerges that the CCM has problems relating to the
observed data.  In Sect.~\ref{sec:tdgs} the problems are interpreted
as clues to a possible solution of the origin of the satellite
galaxies.  The implications of testing the CCM on the Local Group for
gravitational theories are also discussed.  Conclusions regarding the
consequences of this are drawn in Sect.~\ref{sec:concs}.

\section{The satellite mass -- luminosity relation (problem~1)}
\label{sec:ML}

Our understanding of the physical world relies on some fundamental
physical principles. Among them is the conservation of energy. This
concept implies that it is increasingly more difficult to unbind
sub-components from a host system with increasing host binding energy.

Within the DM hypothesis, the principle of energy conservation
therefore governs how DM potentials fill-up with matter. There are two
broadly different physical models exploring the consequences of this,
namely models of DM halos based on internal energy sources (mostly
stellar feedback), and models based on external energy input (mostly
ionisation radiation). In the following, the observational
mass--luminosity data for the known satellite galaxies are discussed,
and the data are then compared to the theoretical results that are
calculated within the CCM.

\subsection{The observational data}
\label{ssec:obsLM}

Based on high quality measurements of individual stellar line-of-sight
velocities in the satellite galaxies, \cite{Strigari08} (hereinafter
S08) calculate dynamical masses, $M_{\rm 0.3kpc}$, within the
inner~0.3~kpc of~18 MW~dSph satellite galaxies over a wide range of
luminosities ($10^3 \simless L/L_\odot \simless 10^7$). The LMC and
SMC are excluded, as is Sagittarius because it is currently
experiencing significant tidal disturbance.  S08 significantly
improve the previous works by using larger stellar data sets and 
more than double the number of dwarf galaxies, and by applying more
detailed mass modelling. Their results confirm the earlier suggestion
by \cite{Mateoetal93}, \cite{Mateo98}, \cite{Giletal07}, 
and \cite{ Penarrubia08}
that the satellites share a common DM mass scale of about
$10^7\,M_\odot$, ``and conclusively establish'' (S08) this common mass
scale.

The finding of S08 can be quantified by writing
\begin{equation}
{\rm log}_{10}M_{\rm 0.3 kpc}= {\rm log}_{10}M_0 + \kappa\,{\rm log}_{10}L,
\label{eq:ML}
\end{equation}
and by evaluating the slope, $\kappa$, and the scaling, $M_0$. 
 S08 derive 
 $\kappa=0.03\pm0.03$ and $M_0 \approx 10^7\,M_\odot$.  Using
the Dexter Java application of \cite{Demleitneretal01}, a nonlinear,
asymmetric error weighted least squares fit to the S08 measurements
reproduces the common mass and slope found by S08, as can be seen from
the parameters listed in Table~\ref{tab:fits}.  By excluding the
least luminous dSph data point, one obtains the same result
(Table~\ref{tab:fits}).

It follows from Eq.~\ref{eq:ML} that
\begin{eqnarray}
(M_{\rm 0.3 kpc})^{1/\kappa} & = & M_0^{1/\kappa}\,L \quad (\kappa \ne 0), \nonumber \\
M_{\rm 0.3 kpc}            & =  & M_0             \quad\quad\; (\kappa=0).
\label{eq:exp}
\end{eqnarray}
This central mass of the DM halo can be tied by means of high-resolution
CDM simulations to the total halo virial mass before its fall into the
host halo (S08, see also Sect.~\ref{sec:mfn}),
\begin{equation}\label{eqn:Mh=M0.3}
M_{\rm vir} = (M_{\rm 0.3 kpc})^{1/0.35}\times10^{-11}M_\odot,
\label{eq:bullock}
\end{equation}
yielding $M_{\rm vir}=10^9\,M_\odot$ for $M_{\rm 0.3
  kpc}=10^7\,M_\odot$ (the common-mass scale for $\kappa=0$). Thus,
substituting $M_{\rm 0.3 kpc}$ into Eq.~\ref{eq:bullock} using
Eq.~\ref{eq:exp} with $\kappa\ne 0$, leads to
\begin{equation}\label{eqn:MhLv}
(M_{\rm vir})^{0.35/\kappa} = M_0^{1/\kappa}\times10^{-(11\times0.35)/\kappa}\,L.
\label{eq:etakappa}
\end{equation}
This value of the halo mass near $10^9\,M_\odot$ for the satellites in
the S08 sample is confirmed by a new analysis, in which
\cite{Wolfetal09} show that the mass can be derived from a velocity
dispersion profile within the deprojected 3D half light profile with
minimal assumptions about the velocity anisotropy. In this way they
obtain a robust mass estimator.

The observed 5$\sigma$ lower value for $0.35/\kappa \equiv \eta$ is
thus~$\eta = 2.06$ (with $\kappa=0.02+5\times0.03$ from
Table~\ref{tab:fits}).

\subsection{Model type A: Internal energy sources}

\cite{DekelSilk86} and \cite{DekelWoo03} studied models according to
which star formation in DM halos below a total halo mass of $M_{\rm
  vir} \approx 10^{12}M_\odot$ is governed by the thermal properties of
the inflowing gas, which is regulated primarily by supernova feedback. 
These models demonstrate that the mass-to-light ratio of sub-halos
follows $M_{\rm vir}/L \propto L^{-2/5}$ (eq.~24 of
\citealt{DekelWoo03}; see also eq.~33 of \citealt{DekelSilk86}). This
approximately fits the observed trend for dSph satellite galaxies
\citep{Mateo98}.

These models thus imply that 
\begin{equation}\label{eq:DS}
\left(M_{\rm vir}\right)^{\eta_{\rm th}} = \zeta \; L,
\end{equation}
where $L$ is the total luminosity, $M_{\rm vir}$ is the virial DM halo
mass, $\eta_{\rm th}=5/3$, and $\zeta$ is a proportionality factor. In
essence, this relation states that more-massive
halos have a larger binding energy such that it becomes more difficult
to remove matter from them than from less massive halos.

Comparing with Eq.~\ref{eq:etakappa} and with its resulting $\eta$
value as given at the end of Sect.~\ref{ssec:obsLM}, it follows that
the observed 5$\sigma$ lower value for $\eta = 0.35/\kappa =2.06$
is in conflict with Eq.~\ref{eq:DS} where $\eta_{\rm th}=5/3=1.67$.

\begin{table}
\begin{center}
\caption[]{The slope of the DM-mass--luminosity relation of dSph
  satellite galaxies.  Fitted parameters for
  Eq.~\ref{eq:ML}.\label{tab:fits}}
\begin{tabular}{lccc}
\hline\hline\\[-3mm]
data   &$\kappa$ &radius  &$M_0$\\
	  &		    & [pc]	 & $[10^7\,M_\odot]$\\
\hline

{\bf Observational:}\\

1   & $+0.02\pm0.03$  &300
&$1.02\pm0.39$\\

2 & $+0.02\pm0.03$  &300   
&$1.01\pm0.40$\\

3 & $+0.01\pm0.03$  &300   
&$1.09\pm0.44$\\

*4 & $-0.03\pm0.05$  &600   
&$6.9\pm4.9$\\

\hline

{\bf DM Models:}\\

A:$\;$ feedback &$0.21$  &300
&---\\

B1:$\;$ re-ionisation, SPS &$0.15\pm0.02$  &300
&$0.24\pm0.06$\\

B2:$\;$ re-ionisation &$0.17\pm0.01$  &300
&$0.18\pm0.02$\\

C:$\;$ SAM &$0.42\pm0.02$  &300
&$2.0\pm0.9$\\

*D:$\;$ Aq-D-HR &$0.17\pm0.02$  &600
&$0.41\pm0.14$\\

E1:$\;$ 1keV(WDM) &$0.23\pm0.04$  &300
&$0.069\pm0.045$\\

E2:$\;$ 5keV(WDM) &$0.12\pm0.02$  &300
&$0.43\pm0.081$\\

F:$\;$ Aq-infall &$0.13\pm0.01$  &300
&$0.32\pm0.022$\\

\hline
\end{tabular}
\end{center}
Notes to the table: Fits to $\kappa=0.35/\eta$: data 1--4 are
observational values, data A--F are models (see Sect.~\ref{sec:ML}).
Notes: 1: our fit to S08 (who give central 300~pc masses, 18
satellites, their fig.~1). 2: our fit to S08 without Seg.1 (faintest
satellite, i.e. 17 satellites, their fig.~1). 3: our fit to S08
without Seg.1 and without Hercules (i.e. 16 satellites, their fig.~1).
4: our fit to the observational data plotted by \cite{OF09} (who give
central 600~pc masses, only 8 satellites, their fig.~1). 
A: \cite{DekelSilk86,DekelWoo03}, stellar feedback (Eq.~\ref{eq:DS}). 
B1: our fit to \cite{Busha09}, their SPS model. B2: our fit to
\cite{Busha09}, inhomogeneous re-ionisation model. C: our fit to
\cite{Maccioetal09}, semi-analytical modelling (SAM), fit is for
$L_V>3\times10^5\,L_{V,\odot}$. D: our fit to \cite{OF09} (Aq-D-HR).
E1: our fit to the $1\,$keV WDM model of \cite{MF09}. E2: our fit to
the $5\,$keV WDM model of \cite{MF09}. F: our fit to the Aquarius
sub-halo-infall models of \cite{Cooper10}. *: the entries with an
asterisk are for the central 600~pc radius region.
\end{table}

\subsection{Model type B1, B2: External energy source}

\cite{Busha09} follow a different line of argument to explain the
dSph satellite population by employing the DM halo distribution from
the {\sl via Lactea} simulation.  Here the notion is that
re-ionisation would have affected DM halos variably, because of an
inhomogeneous matter distribution. A given DM halo must grow above a
critical mass before re-ionisation to form stars or accrete
baryons. Thus the inhomogeneous re-ionisation model
(\citealt{Busha09}{, their fig.~6}) implies, upon extraction of the
theoretical data and using the same fitting method as above,
theoretical $\kappa$-values of 0.15--0.17. These disagree however,
with the observational value of~0.02 with a significance of more than
4~$\sigma$, i.e. the hypothesis that the observational data are
consistent with the models can be discarded with a confidence of
99.99~per cent (Table~\ref{tab:fits}).

\cite{Busha09} suggest that adding scatter into the theoretical
description of how DM halos are filled with luminous baryons would
reduce the discrepancy, but it is difficult to see how this can be
done without violating the actual scatter in the observed $M_{\rm 0.3
  kpc}-L$ relation. 

\subsection{Model type C: Semi-analytical modelling (SAM)}

Filling the multitude of DM halos with baryons given the above
combined processes was investigated by \cite{Maccioetal09}. 
They semi-analytically modelled (SAM) DM sub-halos based on $N-$body
merger tree calculations and high-resolution recomputations. 
The authors state ``We conclude that the number and luminosity of Milky
Way satellites can be naturally accounted for within the
($\Lambda$)Cold Dark Matter paradigm, and this should no longer be
considered a problem.''

Their theoretical mass--luminosity data are plotted in their fig.~5,
and a fit to the redshift $z=0$ data for
$L_V>3\times10^5\,L_{V,\odot}$ satellites is listed in
Table~\ref{tab:fits}. The theoretical SAM data set shows a steep
behaviour, $\kappa=0.42$.  Given the observational data, this model is
ruled out with a confidence of more than ten~$\sigma$.

\subsection{Model type D: High-resolution baryonic physics simulations (Aq-D-HR)}

The satellite population formed in a high-resolution N-body
$\Lambda$CDM re-simulation with baryonic physics of one of the MW-type
``Aquarius'' halos is studied by \cite{OF09}. The treatment of baryonic
processes include time-evolving photoionisation, metallicity-dependent
gas cooling and photo-heating, supernova (SN) feedback, and
chemical enrichment by means of SN~Ia and~II and AGB stars. Re-ionisation
is included and the galactic winds driven by stellar feedback are
assumed to have velocities proportional to the local velocity
dispersion of the dark-matter halo. In these models 100~per cent of the
SNII energy is deposited as thermal energy.  Galactic winds are thus
produced even for the least-massive dwarf galaxies. Winds are observed
in strong starbursts induced through interactions rather than in
self-regulated dwarf galaxies, which may pose a problem for this ansatz
\citep{Ott05}.  The details of the simulations are provided by
\cite{Okamotoetal09}. The resultant sub-halo population with stars
can, as claimed by the authors, reproduce the S08 common-mass scale.

Following the same procedure as for the above models, this claim is
tested by obtaining $\kappa$ from their fig.~1 (upper panel, red
asterisks) and comparing it to the observational data also plotted in
their fig.~1 (note that \citealt{OF09} plot the masses within 600~pc
rather than 300~pc as used above). 
 From their plot of the observational data,
which only includes central-600~pc masses for
the eight most luminous satellites, it follows 
that $\kappa_{\rm obs, OF} =
-0.03\pm0.05$. This is nicely consistent with the full S08 sample
(18~satellites) discussed above.  However, for their model data one
finds that $\kappa=0.17\pm0.02$, i.e. the model can be discarded with
a confidence of 3$\sigma$ or 99.7~per cent.

\subsection{Model type E1, E2: WDM}
\label{ssec:WDM}

\cite{MF09} present theoretical distributions of satellite galaxies
around a MW-type host halo for different cosmological models, namely
$\Lambda$CDM and WDM with three possible DM-particle masses of
$m_{\rm WDM}=1$, 2, and 5~keV. They perform numerical structure formation
simulations and apply semi-analytic modelling to associate the DM
sub-halos with luminous satellites.  
They suggest the luminosity function and
mass--luminosity data of observed satellites is 
reproduced by the WDM models implying a possible lower limit to the
WDM particle of $m_{\rm WDM}\approx 1\,$keV.

The model and observational mass--luminosity data are compared in
their fig.~5 for $m_{\rm WDM}=1$ and~5~keV. The slopes of these model data are
listed in Table~\ref{tab:fits}. From Table~\ref{tab:fits} it follows
that the WDM model with $m_{\rm WDM} \approx 1$~keV is ruled out with very
high confidence (4$\sigma$ or 99.99~per cent), and also has too few
satellites fainter than $M_V\approx -8$ (their fig.~4).  WDM models
with $m_{\rm WDM}\approx 5$~keV are excluded at least with a 3$\sigma$ or
99.7~per cent confidence, and, as is evident from their fig.~4, the
models contain significantly too few satellites brighter than $M_V =-11$.

\subsection{Model type F: Infalling and disrupting dark-matter satellite galaxies}

\cite{Cooper10} study CDM model satellites in individual numerical
models of dark matter halos computed within the Aquarius
project. Semi-analytical modelling is employed to fill the sub-halos
with visible matter, and the orbits of the infalling satellites are followed. 
General agreement with the observed satellites is claimed. 

Much as the other models above, in this numerical CDM model of substructure
and satellite formation in a MW type host halo, the MW sub-halos
fall-in stochastically and therefore do not agree with the observed
phase-space correlated satellites, i.e. with the existence of a
rotating DoS (Sect.~\ref{sec:DoS} below).  Furthermore, the presented
model mass-luminosity data (their fig.~5) lead to a too steep slope
(Table~\ref{tab:fits}) compared to the observations and the DM-based
model is excluded with a confidence of at least 99.7~per cent.
In addition, fig.~5 of \cite{Cooper10} shows a significant increase
in the number of model satellites with a similar brightness as the
faintest known satellite (Segue~1, hereinafter Seg.\,1).
This is in contradiction with the
failure to find any additional satellites of this luminosity in the
most recent data mining of the existing northern SDSS data, as
discussed in Sect.~\ref{ssec:sub} below. Indeed, observations suggest
that Seg.~1 is a star cluster rather than a satellite galaxy
\citep{Niederste09}, worsening this problem.

\subsection {Discussion}
\label{ssec:ABC}

\begin{figure}
\includegraphics[angle=0,scale=0.43]{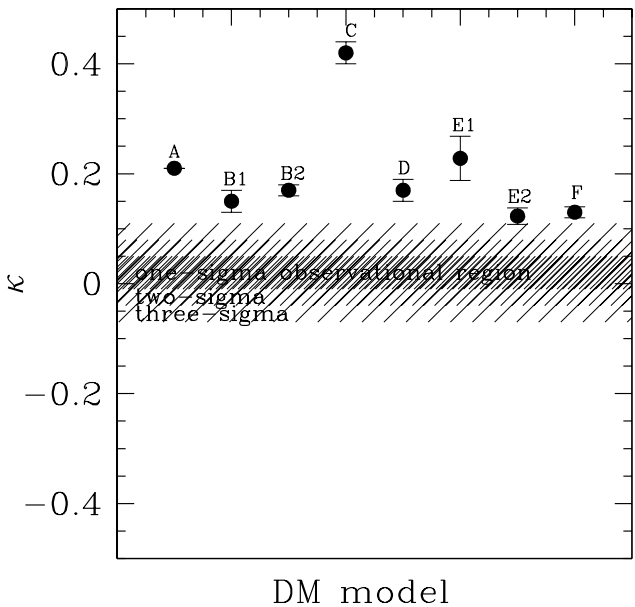}
\vspace{0mm}
\caption{The slope of the mass--luminosity relation, $\kappa$
  (Eq.~\ref{eq:ML}), for the models listed in
  Table~\ref{tab:fits}. The observational constraints with confidence
  intervals are depicted as hatched regions (1, 2, and
  3$\sigma$ region).  Satellites with a larger dark-matter mass are
  on average more luminous such that the mass--luminosity relation has
  $\kappa>0$. However, the observational constraints lie in the region
  $\kappa\approx 0$ (see Table~\ref{tab:fits}). The hypothesis that
  the data are consistent with any one of the models can be discarded
  with very high (at least 3$\sigma$, or more than 99.7~per cent)
  confidence.
\label{fig:fits}}
\end{figure}

In Fig.~\ref{fig:fits}, the latest theoretical ansatzes~A--F to solve
the cosmological substructure problem are compared with the latest
observational limit on the slope $\kappa$ of the DM-mass--luminosity
relation of dSph satellite galaxies (Eq.~\ref{eq:ML}).

The theoretical results always lead to a trend of luminosity with halo
mass as a result of energy conservation. But the observed satellites
do not show an increasing trend of luminosity with DM mass, according
to \cite{Mateo98}, \cite{Penarrubia08}, and \cite{Strigari08}. From
Fig.~\ref{fig:fits} we note that seven $\Lambda$CDM models of the
satellites deviate $4 \sigma$ or more from the data, while only one
(the WDM model E2 with $m_{\rm WDM}=5\,$keV, Table~\ref{tab:fits})
deviates more than $3 \sigma$ from the data. The likelihood
\footnote{The {\sl likelihood} $=$$1-$(confidence in per cent)/100 gives
an indication of how well the data can be accounted for by a given model.
 The {\sl confidence}, as used throughout this text, is the
  probability level at which a model can be discarded.} that any of
the DM models describes the data is thus less than 0.3~per cent.

As a caveat, the observed absence of a DM-mass-luminosity relation
partially depends on the data for the ultra-faint dwarfs: indeed, for
the classical (most luminous) dSphs, \cite{Serra09} argue that there
may be a trend, $\kappa>0$, essentially because of their proposed
increase in the mass of the Fornax dSph satellite. It is on the other
hand plausible that the ultra-faint dwarfs do not possess any dark
halo (see Sect.~\ref{sec:tdgs}), and that the enclosed mass derived is
due to observational artifacts. In that case they should not be used
as a possible improvement for the missing satellite problem. This,
however, would pose a problem for the DM hypothesis.

\cite{Adenetal09b} suggest that for the Hercules dSph satellite
inter-loper stars need to be removed from the observational sample,
which would require a revision of the mass within 300~pc to the value
$M_{\rm 0.3 kpc}=1.9^{+1.1}_{-1.6}\times 10^6\,M_\odot$ (instead of
the value $M_{\rm 0.3 kpc}=7.2^{+0.51}_{-0.21}\times 10^6\,M_\odot$
derived by S08). This new mass measurement, however, now lies more
than $4\,\sigma$ away from all $\Lambda$CDM-models considered above
(Table~\ref{tab:fits}).  Hercules can thus not be understood in
  terms of a DM-dominated model.  \cite{Adenetal09b} do state that 
DM-free models cannot be excluded (note also Fig.~\ref{fig:hercules}
below), or that Hercules may be experiencing tidal disturbances in its
outer parts. Tidal disturbance, however, would have to be very
significant for its inner structure to be affected, because if one
would require conformity with the theoretical DM-models its $M_{\rm
  0.3 kpc}$ mass would have to have been much higher and similar to
the value derived by S08 ($\approx 10^7\,M_\odot$). Given the current
Galactocentric distance of Hercules of 130~kpc and the result that the
inner region of a satellite is only affected by tides after
significant tidal destruction of its outer parts
\citep{Kazantzidisetal04}, this scenario is physically implausible.
There are therefore three possibilities: (i)~Hercules is a
DM-dominated satellite.  This, however, then implies that no logically
consistent solution within the CDM framework is possible because its
mass--luminosity datum would lie well away from the theoretical
expectation. (ii)~Hercules has no DM.  This implies that it cannot be
used in the mass-luminosity data analysis above and would also imply
there to exist an additional type of DM-free satellites, which,
however, share virtually all observable physical characteristics with
the putatively DM filled satellites.  (iii)~Hercules has been
significantly affected by tides.  This case is physically implausible
because of its large distance, but it would imply that Hercules cannot
be used in the mass-luminosity analysis above (just as Sagittarius is
excluded because of the significant tidal effects it is
experiencing). Omitting Hercules from the data leads to a revised
observational slope $\kappa=0.01\pm0.03$ such that none of the
conclusions reached above about the performance of the DM-models are
affected.

A point of contention for DM models of dSph satellite galaxies is that
the DM halos grow at various rates and are also truncated variously
due to tidal influence.  The highly complex interplay between
dark-matter accretion and orbit-induced accretion truncation leads to
the power-law mass function of DM halos, and at the same time would
imply that the outcome in which all luminous DM sub-halos end up
having the same DM mass were incompatible with the DM-theoretical
expectations (see Sect.~\ref{sec:mfn}).

{\sl Summarirising Sect.~\ref{sec:ML}}, while the theoretical results
always lead to a trend of luminosity with halo mass, the observed
satellites do not show this trend.  The hypothesis that the 
CCM accounts for the data can be discarded with more than 99.7~per
cent significance.

\section{The mass function of CDM halo masses (problem~2)}
\label{sec:mfn}

One of the predictions of the $\Lambda$CDM hypothesis is the
self-similarity of DM-halos down to (at least) the mass range of dwarf
galaxies, i.e. that massive halos contain sub-halos of lower mass,
with the same structure in a statistical sense (\citealt{Mooreetal99};
for a major review see \citealt{DelPopolo2007}). The mass function of
these sub-halos is, up to a critical mass $M_{\rm{crit}}$, well
approximated by
\begin{equation}
\xi_{\rm{sub}}(M_{\rm{vir}})=\frac{dN}{dM_{\rm{vir}}} \propto M_{\rm{vir}}^{-1.9},
\label{eq:subhaloMF}
\end{equation}
where $dN$ is the number of sub-halos in the mass interval
$M_{\rm{vir}}, M_{\rm{vir}}+dM_{\rm{vir}}$
\citep{Gaoetal04}, $M_{\rm{crit}}$ is given by $M_{\rm{vir}} \approx
0.01 M_{\rm{h}}$ with $M_{\rm{h}}$ being the virial mass of the hosting
CDM-halo. The virial mass, $M_{\rm{vir}}$, is defined by
\begin{equation} 
M_{\rm{vir}}=\frac{4 \pi}{3} \Delta_{\rm{vir}} \rho_0 r_{\rm{vir}}^3,
\label{eq:VirMass}
\end{equation}
where $\rho_0$ is the critical density of the Universe and
$\Delta_{\rm{vir}}$ is a factor such that $\Delta_{\rm{vir}} \rho_0$
is the critical density at which matter collapses into a virialised
halo, despite the overall expansion of the Universe. The virial radius
$r_{\rm vir}$ is thereby determined by the density profile of the
collapsed CDM-halo. For $M_{\rm{vir}} > 0.01\, M_{\rm{h}}$, the mass
function steepens \citep{Gaoetal04}, so that it is effectively cut off
at a mass $M_{\rm max}$ (see Eq.~\ref{eq:lumMF} below). It is
reasonable to identify $M_{\rm max}$ with the mass of the most massive
sub-halo, which must be higher than $M_{\rm crit}$, where the mass
function begins to deviate from Eq.~\ref{eq:subhaloMF} and lower than
$M_h$, the mass of the host-halo.  Therefore, $M_{\rm crit} < M_{\rm
  max} < M_h$.

Thus, a halo with $M_{\rm{vir}}\approx 10^{12}\, M_{\odot}$, like the
one that is thought to be the host of the MW, should have a population
of sub-halos spanning several orders of magnitude in mass. It is well
known that, in consequence, a steep sub-halo mass function such as
Eq.~\ref{eq:subhaloMF} predicts many more low-mass sub-halos than the
number of observed faint MW satellites \citep{Mooreetal99,Klypin99}, a
finding commonly referred to as the {\sl missing satellite
  problem}. Efforts to solve this problem rely on physical processes
that can either clear CDM-halos of all baryons or inhibit their
gathering in the first place, which would affect low-mass
halos preferentially (e.g. \citealt{Moore06,Lietal09};
Sect.~\ref{sec:ML}). More specifically, \citet{Lietal09} find that the
mass function of luminous halos, $\xi_{\rm{lum}}(M_{\rm{vir}})$, would
essentially be flat for $10^7 M_{\odot} \le M_{\rm{vir}} < 10^9
M_{\odot}$. All sub-halos with $M_{\rm{vir}} \ge 10^9 M_{\odot}$ would
keep baryons and therefore
$\xi_{\rm{lum}}(M_{\rm{vir}})=\xi_{\rm{sub}}(M_{\rm{vir}})$ in this
mass range. Thus, the mass function of {\sl luminous sub-halos} can be
written as
\begin{equation}
\xi_{\rm{lum}}(M_{\rm{vir}}) =k k_i M_{\rm{vir}}^{-\alpha_i},
\label{eq:lumMF}
\end{equation}
with 
 
\begin{math}
\begin{array}{@{\hspace{-0.6cm}}lll}
&&\\[-4pt]
\alpha_1 = 0, & \ k_1=1, & \ 10^7 \le \frac{M_{\rm{vir}}}{M_{\odot}} <
10^9,\\[3pt]
\alpha_2 = 1.9, & \  k_2=k_1\, (10^9)^{\alpha_2-\alpha_1}, & \  10^9 \le  
\frac{M_{\rm{vir}}}{M_{\odot}} \le M_{\rm{max}},\\[-4pt]
&&\\
\end{array}
\end{math}\\
where the factors $k_i$ ensure that $\xi_{\rm{vir}}(M_{\rm{vir}})$ is
continuous where the power changes and $k$ is a normalisation constant
chosen such that
\begin{equation}
\int^{M_{\rm{max}}}_{10^7}\xi_{\rm{vir}}(M_{\rm{vir}}) \, dM_{\rm{vir}}=1.
\label{eq:norm}
\end{equation}
From a mathematical point of view, Eq.~\ref{eq:lumMF} is the
probability distribution of luminous sub-halos.  We note that the
luminous sub-halo mass function proposed in \citet{Moore06} is similar
to the one in \citet{Lietal09}. In the high-mass part, it has the same
slope as the mass function for all sub-halos and flattens in the
low-mass part (cf. fig.~3 in \citealt{Moore06}). The lower mass limit
for luminous halos is however suggested to be $M_{\rm vir} \approx
10^8\,M_{\odot}$ in \citet{Moore06}. The mass function of {\sl all
  sub-halos} has $\alpha_1\approx\alpha_2\approx 1.9$
\citep{Gaoetal04}.

\subsection{NFW halos}
\label{ssec:NFW}

It is well established that the theoretical density profiles of
galaxy-sized CDM-halos are similar to a universal law, as proposed by
\cite{NFW}. The NFW profile is given by
\begin{equation}
\label{eq:NFW}
\rho_{\rm{NFW}}(r) =\frac{\delta _c\rho _0}{r/r_{\rm{s}}\left(
  1+r/r_{\rm{s}}\right) ^2},
\end{equation}
where $r$ is the distance from the centre of the halo and $\rho_0$ is
the critical density of the Universe, while the characteristic radius
$r_{\rm{s}}$ and $\delta_c$ are mass-dependent parameters.

By integrating $\rho_{\rm{NFW}}(r)$ over a volume, the total mass of
CDM within this volume is obtained. Thus,
\begin{equation}
M(r)=\int^r_0 \rho (r')4\pi r'^2 \,dr'
\label{eq:enclmass1}
\end{equation}
is the mass of CDM contained within a sphere of radius $r$ around
the centre of the CDM-halo, and $M(r)=M_{\rm{vir}}$ for
$r=r_{\rm{vir}}$. Performing the integration on the right-hand side of
Eq.~\ref{eq:enclmass1} and introducing the concentration parameter
$c=r_{\rm{vir}}/r_{\rm{s}}$ leads to
\begin{equation}
M(r) =\frac{4 \pi \rho_0 \delta _c r_{\rm vir}^3}{c^3} \; \left[
  \frac{r_{\rm{vir}}}{r_{\rm{vir}}+c\;r}+\ln \left(
  1+\frac{c\;r}{r_{\rm{vir}}} \right) -1\right].
\label{eq:enclmass2}
\end{equation}
The parameter $\delta_c$ can be expressed in terms of $c$, 
\begin{equation}
\delta _c =\frac{\Delta_{\rm{vir}}}{3} \frac{c^3}{\ln \left( 1+c
  \right) -c/(1+c)},
\label {eq:delta}
\end{equation}
as can be verified by setting $r=r_{\rm{vir}}$ in
Eq.~\ref{eq:enclmass2} and substituting $M(r_{\rm vir})=M_{\rm vir}$
by Eq.~\ref{eq:VirMass}.

If the halo is luminous, it is evident that $M(r)$ is smaller than the
total mass included within $r$, $M_r$. However, assuming that the MW
satellites are in virial equilibrium and that their dynamics is
Newtonian in the weak-field limit, the mass-to-light ratios calculated
for them are generally high and imply that they are DM-dominated and
thus, $M(r)=M_r$ would be a good approximation. This relation is
therefore adopted for the present discussion. In this approximation
$M(r=0.3 {\rm kpc})=M_{\rm 0.3kpc}$.

In principle, the parameters $\rho_0$ \citep{NFW}, $c$
\citep{Bullock01}, and $\Delta_{\rm{vir}}$ \citep{Mainini03} depend on
the redshift $z$ but for the purpose of the present paper only $z=0$
needs to be considered, as this is valid for the local Universe. Thus,
\begin{equation}
\rho_0 =\frac{3H_0^2}{8\pi G},
\label{rhocrit}
\end{equation}
where the Hubble constant $H_0=71 \, \rm{km} \, \rm{s}^{-1} \,
\rm{Mpc}^{-1}$ \citep{Spergel07}, $\Delta_{\rm{vir}} \simeq 98$ for
$\Lambda$CDM-cosmology \citep{Mainini03}, and
\begin{equation}
\log_{10}(\overline{c})=2.31-0.109
\log_{10}\left(\frac{M_{\rm{vir}}}{M_{\odot}}\right),
\label{eq:c}
\end{equation}
where $\overline{c}$ is the expectation value of $c$ as a function of
$M_{\rm{vir}}$. Thus, $\overline{c}$ decreases slowly with
$M_{\rm{vir}}$, while the scatter in the actual $c$ is rather large,
being
\begin{equation}
\sigma_{\log_{10} c}=0.174
\label{eq:sigc}
\end{equation}
\citep{Maccio07}. The only caveat here is that the NFW
profile is used to integrate the mass, while the now-preferred Einasto
profile (\citealt{Navarro09}, Sect.~\ref{sec:introd}) makes only a
small difference in the central parts.

\subsection{Probing the $\Lambda$CDM hypothesis with $M_{\rm 0.3kpc}$}

\begin{figure}
\includegraphics[angle=0,scale=0.80]{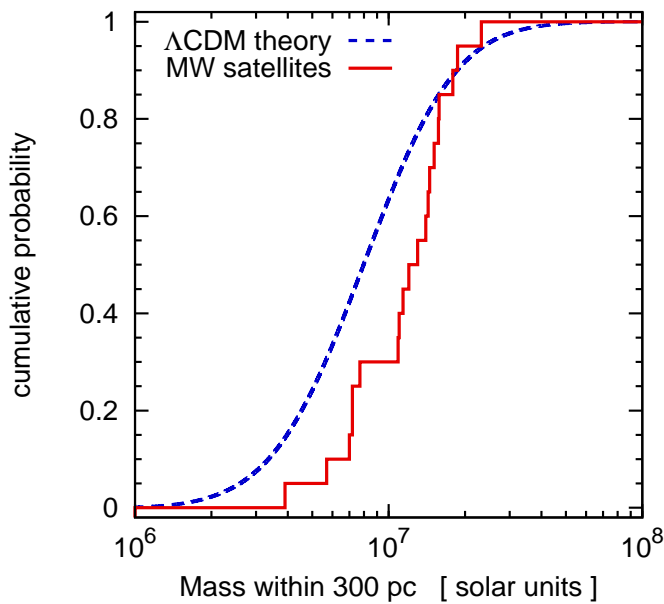}
\vspace{0mm}
\caption{The {\sl mass function of luminous satellite problem}. The
  cumulative distribution function for the mass within the central
  300~pc, $M_{\rm 0.3kpc}$, of the MW satellites (solid line) and the
  cumulative distribution function for $M_{\rm 0.3kpc}$ of a sample of
  $10^6$ CDM-halos picked from the parent distribution of luminous
  sub-halos (Eq.~\ref{eq:lumMF}, dashed line). The null hypothesis is
  that the MW satellite $M_{\rm 0.3 kpc}$ masses are drawn from this
  parent distribution.  The maximum distance between the two curves is
  0.333 so that the null hypothesis can be discarded with~98.9~per
  cent confidence.}
\label{fig:DMscatter}
\end{figure}

S08 use the stellar motions in 18 MW satellites to calcule their mass
within the central 300 pc, $M_{\rm 0.3kpc}$. They assume the
satellites to be in virial equilibrium and that Newtonian dynamics can
be applied to them. The sample from S08 can be enlarged to
20~satellites by including the Large Magellanic Cloud (LMC) and the
Small Magellanic Cloud (SMC), since \citet{vanderMarel02} estimated
the mass of the LMC within the innermost 8.9~kpc, $M_{\rm{LMC}}$,
using the same assumptions as S08.  This implies that
$M_{\rm{LMC}}=(8.7\pm4.3)\times 10^9 \, M_{\odot}$, of which the major
part would have to be DM.
Equations.~\ref{eq:VirMass},~\ref{eq:enclmass2},~\ref{eq:delta}
and~\ref{eq:c} have been used to create tabulated expectation values
of $M(r)$ for NFW-halos with different $M_{\rm{vir}}$ and it can
thereby be seen that for a typical NFW-halo with $M(r=8.9 \, {\rm
  kpc})=8.7 \times 10^9 \, M_\odot$, $M(r=0.3 \, {\rm kpc})=2.13
\times 10^7 \, M_\odot = M_{\rm 0.3kpc}$, and $M_{\rm vir}=1.2\times
10^{11} \, M_\odot$. We note that the SMC has about 1/10th of the mass
of the LMC \citep{Kallivayalil06}, hence the virial mass of its halo can be
estimated as $M_{\rm vir}=1.2\times 10^{10} \, M_\odot$, corresponding
to $M_{\rm 0.3kpc}=1.51 \times 10^7 \, M_\odot$.

To test the shape of the MW satellite distribution function
against the shape of the distribution of the $M_{\rm 0.3kpc}$ values
of the MW-satellites, artificial samples of $10^6 \ M_{\rm 0.3kpc}$
masses are generated in concordance with the $\Lambda$CDM hypothesis,
using Monte Carlo simulations. As noted in Sect.~\ref{ssec:NFW},
$M_{\rm 0.3kpc}$ is well approximated by $M(r=0.3 \rm{kpc})$ in a
CDM-dominated galaxy. $M(r=0.3 \rm{kpc})$ can be calculated if
$M_{\rm{vir}}$ and $c$ are given, and the expectation value for $c$ is
a function of $M_{\rm{vir}}$. The first step is therefore to choose a
value for $M_{\rm{vir}}$ using uniform random deviates and the
probability distribution of luminous halos given in Eq.~\ref{eq:lumMF}
(see e.g. chapter~7.2 in \citealt{NumRecipes} for details). The next
step is to attribute a value for $\log_{10}(c)$ to the chosen
$M_{\rm{vir}}$. This is done by multiplying Eq.~\ref{eq:sigc} with a
Gaussian random deviate and adding the result to the value for
$\log_{10}(\overline{c})$, which is calculated from
Eq.~\ref{eq:c}. After transforming $\log_{10}(c)$ to $c$, $M_{\rm
  0.3kpc}=M(r=0.3 \rm{kpc})$ of the given halo can be calculated from
Eq.~\ref{eq:enclmass2}, using Eqs.~\ref{eq:VirMass} and~\ref{eq:delta}.
These steps are repeated, until a sample of $10^6 \ M_{\rm 0.3kpc}$
values is generated.

If two samples are given, the maximum distance between their
cumulative distribution functions, $D$, can be calculated.  Performing
the KS-test, this quantity $D$ allows an estimate of how likely it is that
they are drawn from the same distribution function. The null
hypothesis is that the observed satellite galaxies are drawn from the
theoretically calculated mass function of luminous halos; the parent
distribution is thus assumed to be the mass function of $M({\rm 0.3
  kpc})$ values of luminous sub-halos according to the $\Lambda$CDM
hypothesis. 
Assuming in Eq.~\ref{eq:lumMF} that $M_{\rm{max}} = 10^{11} M_{\odot}$,
which is approximately the mass estimated for the CDM halo
of the LMC, and taking $M_{\rm min}=10^7\,M_\odot$, leads to
$D=0.333$.  According to the KS-test, given the parent distribution
the probability of an even larger distance is~0.011.  This means that
the null hypothesis can be excluded with 98.9~per cent
confidence. Both cumulative distributions are shown in
Fig.~\ref{fig:DMscatter}\footnote{ Monte Carlo experiments are used to
  quantify the confidence values for the KS-tests: Drawing the
  corresponding number of sub-halo masses (e.g.~20 as in this case)
  from Eq.~\ref{eq:lumMF}, $D'$ is calculated. This is repeated $10^5$
  times. Counting of $D'$ values gives the fraction of cases when
  $D'>D$, where $D$ is the actually obtained $D'$ value from the data
  (e.g. $D=0.333$ in this case). These fractions are reported here as
  likelihood values, and are about half as large as the probability
  values obtained using approximate methods, as, e.g., by 
  \cite{NumRecipes}.}.

Omitting the LMC and SMC from the observational sample but keeping
$M_{\rm min}=10^7\,M_\odot$ and $M_{\rm max}=10^{11}\,M_\odot$ in the
theoretical sample yields $D=0.294$, leading to the exclusion of the null
hypothesis with a confidence of 95.5~per cent.  In addition setting
$M_{\rm max} = 4\times10^{10}\,M_\odot$, which is the $M_{\rm vir}$
that corresponds to the most massive $M_{\rm 0.3 kpc}$ in the S08
sample (i.e. the most massive remaining sub-halo), yields $D=0.301$
leading to exclusion of the null hypothesis with a confidence of
96.3~per cent.  The latter two tests comprise a homogeneous
mass-sample of observed satellites as compiled by S08.

That the mass function is expected to steepen at
$M_{\rm{crit}}=0.01\,M_{\rm{h}}$ even increases the discrepancy
between the $\Lambda$CDM hypothesis and the observations.  Reinstating 
the LMC and SMC back into the observational sample and cutting off
$\xi_{\rm{sub}}(M_{\rm{vir}})$ at $M_{\rm{max}}=10^{10} M_{\odot}$
(with $M_{\rm min}=10^7\,M_\odot$), which would be close to
$M_{\rm{crit}}$ for the CDM-halo of the MW (see Sect.~\ref{sec:mfn}),
and one order of magnitude below the estimated mass of the CDM-halo of
the LMC, implies that $D=0.359$ and an exclusion with 99.5~per cent
confidence.  

On the other hand, setting $M_{\rm{max}}=10^{12} \, M_{\odot}$ (with
$M_{\rm{min}}=10^7 \, M_{\odot}$) leads to $D=0.329$ and an exclusion
with 98.8 per cent confidence. Any reasonable uncertainty in the
actual value of $M_{\rm{max}}$ can therefore be excluded as an
explanation of the discrepancy between the observed sample of
$M_{0.3\, \rm{kpc}}$ and a sample generated based on the $\Lambda$CDM
hypothesis. As a consequence, the same is true for the uncertainty in
the actual mass of the halo of the MW, $M_{\rm{h}}$, since
$M_{\rm{max}}$ is linked to $M_{\rm{h}}$ (see Sect.~\ref{sec:mfn}).

Thus $M_{\rm{max}}$ is kept at $10^{11} \, M_{\odot}$ in the
following. 
Adjusting the lower limit of $\xi_{\rm{lum}}(M_{\rm{vir}})$
 from $10^7 \, M_{\odot}$ to $10^8 \, M_{\odot}$ then leads to
$D=0.319$ and an exclusion of the null-hypothesis with a confidence of
98.4 per cent. The mass of $10^8 \, M_{\odot}$ is the $M_{\rm{vir}}$
suggested by the lowest $M_{0.3\rm{kpc}}$ in the sample from S08.
We note that the likelihood decreases with decreasing $M_{\rm max}$, 
because of the overabundance of $M_{\rm 0.3\,kpc}\approx10^7\,M_\odot$
halos becoming more prominent in the observational sample.

S08 suggest that $\xi_{\rm{lum}}(M_{\rm{vir}})$ might even be cut off
below a mass of $\approx 10^9 M_{\odot}$, either because halos below
that mass do not contain baryons or do not form at all. Indeed,
modifying $\xi_{\rm{lum}}(M_{\rm vir})$ given by Eq.~\ref{eq:lumMF}
accordingly, results in an agreement between the theoretical
distribution and the data ($D=0.188$ with an exclusion confidence of only
70~per cent). A $\xi_{\rm{lum}}(M_{\rm{vir}})$ with a lower mass limit
of $10^9 \, M_{\odot}$ is however in disagreement with the
$\Lambda$CDM hypothesis, since the limiting mass below which all
CDM-halos are dark ought to be two orders of magnitude lower according
to \citet{Lietal09}.

As a final note, the newly derived reduced mass of Hercules
(see end of Sect.~\ref{ssec:ABC}) affects neither the calculated
likelihoods nor the conclusions reached here.

{\sl Summarising Sect.~\ref{sec:mfn}}, the mass distribution of the
predicted DM halos of observed satellites is consistent with the
$\Lambda$CDM hypothesis with at most 4.5~per cent likelihood.
Assuming the dSph satellites are in virial equilibrium, the
observationally deduced DM halo masses of the MW satellites show a
significant overabundance of $M_{\rm 0.3kpc}\approx 10^7\,M_\odot$
halos and a lack of less-massive values compared to the theoretically
calculated distribution for luminous sub-halos, despite much effort to
solve the {\sl common-mass-scale problem} (Sect.~\ref{sec:ML}).

\section{The bulge mass versus satellite number relation (problem~3)}
\label{sec:origin}

According to a straight forward interpretation of the CCM, more
massive DM host halos have a larger number of luminous satellites
because the number of sub-halos above a low-mass threshold increases
with host halo mass, given the host halo mass waxes by accreting
sub-halos.  The sub-halos are accreted mostly individually without a
physical link to the processes occurring at the centre of the host
halo.  There indeed does not appear to be an observed relation
between the halo mass and the bulge mass, since pairs of galaxies with
and without a bulge (such as M31, \citealt{RF70}, and M101,
\citealt{Bosmaetal81}, respectively) but with the same rotation
velocity can be found. It would be useful to return to models A--F
(Sect.~\ref{sec:ML}) and to include the formation of the host galaxy
in the modelling to quantify the degree of correlation between the
bulge mass and number of luminous satellites actually expected in the
CCM. When doing so, the same type of models will also have to account
for the presence of bulge-less galaxies having the same DM-halo mass,
as pointed out above. That is, it would {\sl not} suffice to merely
demonstrate that some sort of bulge-mass--satellite number correlation
emerges in the CCM. The case $M_{\rm bulge}=0$ must emerge naturally
within the model, since two-thirds of all bright disk galaxies have no
bulge or only a small one \citep{C009b}.

On the basis of extragalactic observational data,
\cite{Karachentsevetal05} note, but do not quantify, the existence of
a correlation between the bulge luminosity and the number of
associated satellite galaxies such that galaxies without a bulge have
no known dSph companions, such as M101. 
\cite{Karachentsevetal05} also point out that the number of known dSph
satellites increases with the tidal environment.

\begin{figure}
\includegraphics[angle=0,scale=0.43]{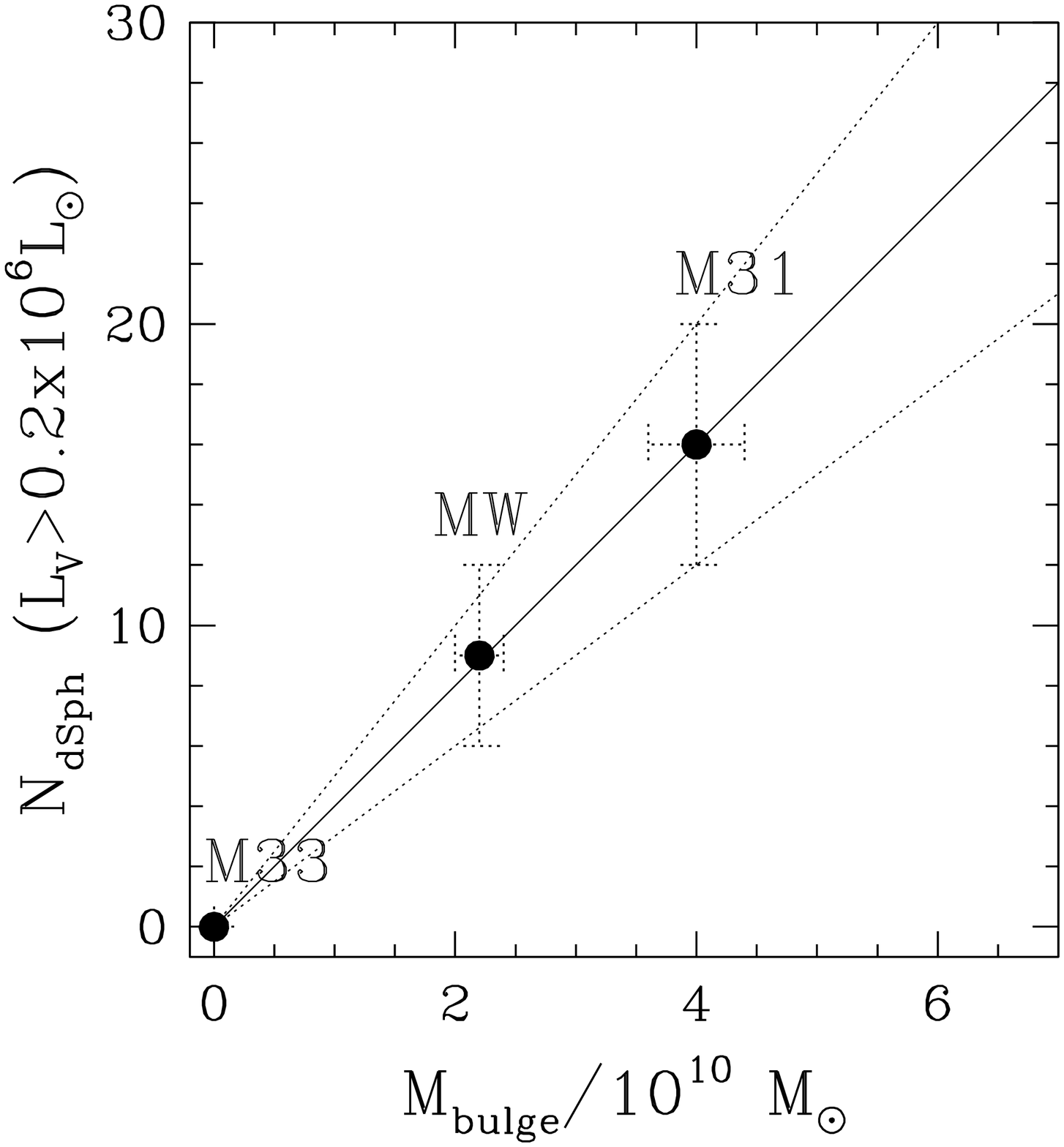}
\vspace{-15mm}
\caption{The number of dSph and dE satellite galaxies more luminous
  than $0.2\times10^6\,L_\odot$ is plotted versus the bulge mass of
  the host galaxy (MW: \citealt{Zhao96}, M31: \citealt{Kent89}, M33:
  \citealt{Gebhardtetal01}). Only satellites within a distance of
  270~kpc of the MW and M31 are used. The solid line (slope$=4.03$) is
  Eq.~\ref{eq:bulge}.  The upper (slope$=5.03$) and the lower
  (slope$=3.03$) dotted lines illustrate the relative uncertainty
  assumed in the Monte Carlo experiment (see Sect.\,\ref{sec:origin}).
\label{fig:correl}}
\end{figure}

The existence of this correlation can be tested in the Local Group,
where good estimates of the number of satellites within the nominal
virial radii of the respective hosts and of the stellar bulge masses
of the three major galaxies (MW, M31, and M33) exist.  Only the
satellites brighter than $L_V = 0.2\times10^6\,L_\odot$ ($M_V<-8.44$)
are considered, given that the census of fainter satellites is
incomplete for the MW (notably in the southern hemisphere), and also
for M31 and M33 given their distances.  By restricting this analysis
to satellites with $L_V>0.2\times10^6\,L_\odot$, the result becomes
robust against the discovery of additional satellites since these
would typically be fainter.  The result is displayed in
Fig.~\ref{fig:correl}: a linear correlation between the bulge mass and
the number of early-type satellites is suggested. An error-weighted
least squares linear fit to the data yields
\begin{equation}
N_{\rm dSph} = (4.03\pm 0.04)\times M_{\rm bulge}/(10^{10}\,M_\odot).
\label{eq:bulge}
\end{equation}
In terms of the present-day stellar mass fraction, the dSph satellites
of the MW add-up to at most a few times $10^7\,M_\odot$, so that they
amount to about 0.15~per cent of the mass of the bulge.  Given that
Eq.~\ref{eq:bulge} is a linear fit to three data points only, it will
be important to check the reality of this correlation by surveying
disc galaxies in the Local Volume with different bulge masses for a
deep and exhaustive sampling of satellite galaxies.

Given the small number of observational data points underlying
Eq.~\ref{eq:bulge}, one should not over-interpret this result, but it
is legitimate to inquire how significant the empirical correlation
between bulge mass and the number of satellites is. In view of the
observation by \cite{Karachentsevetal05} noted above, it may be
indicative of a physical correlation.

The significance of the Local Group bulge--satellite correlation is
evaluated by performing a Monte Carlo experiment, the null
hypothesis of which is that there is no correlation. This hypothesis would
appear to be plausible in the CCM because the number of satellites
depends mostly on the host DM halo mass, while the bulge is produced by
baryonic processes taking place near the center of the host DM halo.
Three pairs of $M_{\rm bulge}$ and $N_{\rm dSph}$ values are chosen
randomly from uniform distributions such that $M_{\rm bulge} \in
[0,4.6\times 10^{10}\,M_\odot]$
 and $N_{\rm dSph} \in [0,28]$\footnote{The upper bounds of the
 intervals are the 3$\sigma$ upper
  values of $M_{\rm bulge}$ and $N_{\rm dSph}$ of M31. The scaling of
  the axes is, however, irrelevant for the results of the Monte Carlo
  experiments, because the aim is to test how likely a correlation
  results, given the null hypothesis.}.
  For each three-point data set,
a linear regression yields a measure of the degree of
correlation. This is performed $10^6$ times.  The following incidences
are counted: 1) the resulting linear relation passes the
 $(M_{\rm bulge}, N_{\rm dSph}) = (0,0)$ point\footnote{The precise
 condition here is as follows: 
 Let there be three Monte Carlo pairs
 $(M_{\rm bulge}, N_{\rm dSph})_i, i=1...3$. A linear regression yields a
  slope and an axis intersection, both with uncertainties expressed as
  $\sigma$ values. If the axis intersection lies within $5\sigma$ of the
  $(0,0)$ point, then this particular set of bulge--satellite pairs is
  counted.  Note that the test does not require the slope to be the
  same as the observed value.}
  {\sl and} the slope of the linear
relation has a relative uncertainty smaller than a given value; and
the second test is 2) the slope of the linear relation has a relative
uncertainty smaller than a given value. The relative uncertainty in
the slope used here is based on the uncertainties in the
data. Applying this relative uncertainty to Eq.~\ref{eq:bulge} leads
to $N_{\rm dSph}\approx (4\pm 1)\times M_{\rm bulge}/(10^{10}\,
M_{\odot})$. Taking the upper and the lower 1$\sigma$ limit of the
slope, this equation thereby passes the lower and the upper 1$\sigma$
values of the data (Fig.~\ref{fig:correl})\footnote{The uncertainty
 in the slope given by Eq.~\ref{eq:bulge}
 is a measure for how close the
  data lie to the straight line fitted to them, i.e. very close in the
  given case.  However, the uncertainties on the data suggests that
  the observed case is rather improbable (although obviously not
  impossible), even if the correlation between $N_{\rm dSph}$ and
  $M_{\rm bulge}$ is real. The uncertainty on the slope stated in
  Eq.~\ref{eq:bulge} would therefore not be a good basis for the test
  performed here.}.

The Monte Carlo result is that case 1) occurs~$44\,000$ times, while
case~2) occurs~$157\,000$ times. Thus, if the correlation evident in
Fig.~\ref{fig:correl} were unphysical, then observing it would have a
likelihood of~$0.044$ and~$0.157$, respectively. Given the data on the
Local Group, the above hypothesis that the bulge mass and number of
satellites are not correlated can therefore be discarded with a
confidence of~95.6 per cent and~84.3 per cent 
in case 1) and case 2), respectively.

{\sl Summarising Sect.~\ref{sec:origin}}, the null hypothesis that the
bulge mass and the number of satellites are independent quantities is
rejected, based on the Local Group data, with a confidence of more
than~95.6~per cent.  With the absence of a DM-mass--luminosity
relation for the observed satellites (Sect.~\ref{sec:ML}), this
suggests that our present understanding of how satellite dwarf
galaxies form and evolve may need revision.  In the formation
modelling of satellite galaxies within the CCM it will therefore be
necessary to include also the formation of the host galaxy, to
quantify the correlation between bulge mass and the number of
satellites within the CCM. It will also be essential to refine this
correlation using deep observational extra galactic surveys.

\section{The disc of satellites (DoS) and invariant 
baryonic galaxies (problems~4 and~5)}
\label{sec:DoS}

The DoS is now addressed in view of the new satellite
galaxies, and in Sect.~\ref{ssec:baryonic_invariant} the issue that
the two major DM halos of the Local Group, which happen to be similar,
are occupied by similar disk galaxies is addressed within the context
of the CCM.

An important constraint for understanding the origin and nature of the
observed satellite galaxies comes from them being significantly
anisotropically distributed about the MW, and possibly also about Andromeda.
The problem of the MW system for the CCM was emphasised by
\cite{KTB05}. They pointed out that the observed satellite system of
the MW was incompatible at the 99.5~per cent confidence level with the
theoretical distribution expected if the satellites were DM sub-halos
tracing an isotropic DM host halo.  Until then, the prediction within
the DM hypothesis was that the distribution of sub-halos ought to be
nearly spherical and tracing the shape of the host DM halo. For
example, \cite{APC04} show a MW-type DM halo to have an infall
asymmetry of only about 15~per cent. The sub-halos enter the host halo
along filaments and then phase-mix and virialise within the growing
host DM halo. Similar sub-halo distributions are obtained in CDM and
WDM models \citep{Knebeetal08}.

The DoS is a pronounced feature of the MW satellite system
\citep{Metz09b}, and a similar structure was reported for the
Andromeda system \citep{KG06} for which, however, the distance
uncertainties are larger and the satellite population is richer and
more complex including dSph, dE, and dIrr galaxies. In the case of the
well-studied MW, the DoS is very pronounced for the classical (11
brightest) satellites including the LMC and SMC. But how are the new
satellites, the ultra-faint ones, distributed?  Much hope for the CCM
rests on the new discoveries being able perhaps to alleviate the DoS
problem.

\cite{Watkins09} and \cite{Belokurov10} reported the
discovery of two new MW satellite galaxies, Pisces~I and~II,
respectively, enlarging the total MW satellite system to
24~satellites.  Pisces~I and~II were found in the southern part of the
SDSS survey area, making them the two first non-classical satellite
galaxies found in the Southern Galactic hemisphere.  Furthermore,
distances to a number of the already known satellite galaxies have
been updated in recent years, most notably the new distance estimate
for Boo II by \cite{Walsh08}, which changes the distance from~60 to
42~kpc. An updated list of all currently known satellites is provided
in Table.~\ref{tab:satellites} upon which the following analysis is
based.

\begin{sidewaystable*}
\begin{center}
\caption[]{
Data for the currently known MW satellites.\label{tab:satellites}
}
%\title{}
\begin{tabular}{lrrrrrrcrrrccc}
\hline\hline\\[-3mm]

Name & $\alpha_{2000}$ & $\delta_{2000}$ & $r_{\rm{helio}}$ &  $l_{\rm{MW}}$ & $b_{\rm{MW}}$ & $r_{\rm{MW}}$ & Ref.    & $v_{\rm{GSR}}^{220}$ & $v_{\rm{GSR}}^{250}$ & $\Delta v$ & Ref.   & $L_{\rm{V}}$     & Ref.  \\

 & [h~m~s] & [$^\circ$~m~s] & [kpc] & [$^{\circ}$] & [$^{\circ}$] & [kpc] &  & [kms$^{-1}$] & [kms$^{-1}$] & [kms$^{-1}$] &  & $[L_{\sun}]$ \\ \hline

{\bf New:}\\

Boo &  14 ~ 00 ~ 05 & +14 ~ 30 ~ 21 & $ 64 \pm 2 $ & 357.9 & 76.5 & 61 & 2; 10; 28; 8    & 94.4 & 94.2 & $ \pm 3.4 $ & 21   & $ (2.6 \pm 0.5) \times 10^{ 4 } $ & 2; 18   \\
                               
Boo II & 13 ~ 58 ~ 05 & +12 ~ 51 ~ 36 & $ 45 \pm 2 $  & 348.6 & 78.9 & 43 & 30; 31; 14    &         &                &  &              & $ (9.2 \pm 5.4) \times 10^{ 2 } $ & 30; 18; 31  \\
                               
CVn & 13 ~ 28 ~ 04 & +33 ~ 33 ~ 27 & $ 214 \pm 9 $ & 84.2 & 80.0 & 213 & 36; 8; 16; 17    & 64.8 & 69,7 & $ \pm 0.6 $ & 13; 29   & $ (2.0 \pm 0.3) \times 10^{ 5 } $ & 36; 18  \\
                               
CVn II & 12 ~ 57 ~ 10 & +34 ~ 19 ~ 20 & $ 154 \pm 5 $ & 129.4 & 81.3 & 155 & 25; 3; 12; 8    & -97,5 & -93,2 & $ \pm 1.2 $ & 29   & $ (7.5 \pm 3.1) \times 10^{ 3 } $ & 25; 3; 18  \\
                               
CBe & 12 ~ 26 ~ 59 & +23 ~ 54 ~ 37 & $ 43 \pm 2 $ & 202.2 & 75.5 & 44 & 3; 8; 23    & 47,4 & 40,4 & $ \pm 0.9 $ & 29   & $ (3.1 \pm 1.1) \times 10^{ 3 } $ & 3; 18; 22  \\
                               
Her & 16 ~ 31 ~ 04 & +12 ~ 47 ~ 24 & $ 135 \pm 4 $ & 31.2 & 38.2 & 129 & 3; 7; 8;  1; 26    & 128,6 & 140,0 & $ \pm 1.1 $ & 29;  1   & $ (2.9 \pm 0.7) \times 10^{ 4 } $ & 3; 18; 26  \\
                               
Leo IV & 11 ~ 32 ~ 58 & -00 ~ 32 ~ 09 & $ 156 \pm 5 $ & 261.1 & 56.3 & 156 & 3; 20    & 10,6 & -6,0 & $ \pm 1.4 $ & 29   & $ (1.3 \pm 0.3) \times 10^{ 4 } $ & 3; 18; 27; 9  \\
                               
Leo V & 11 ~ 31 ~ 09 & +02 ~ 13 ~ 05 & $ 176 \pm 10 $ & 257.9 & 58.3 & 176 & 4; 9    & 58,5 & 42,8 & $ \pm 3.1 $ & 4   & $ (6.4 \pm 2.4) \times 10^{ 3 } $ & 4; 9  \\
                               
Pis I & 23 ~ 40 ~ 00 & -00 ~ 18 ~ 00 & $ 80 \pm 14 $ & 100.2 & -57.8 & 80 & 32; 15    & 42,1 & 58,0 &     & 15   &      &    \\
                               
Pis II & 22 ~ 58 ~ 31 & +05 ~ 57 ~ 09 & $ 182 \pm 36 ^{*}$ & 84.1 & -47.6 & 181 & 6    &                 &                &  &        & $ \sim 8.6   \times 10^{ 3 } $ & 6  \\
                               
\textit{Seg I} & 10 ~ 07 ~ 04 & +16 ~ 04 ~ 40 & $ 23 \pm 2 $  & 206.2 & 39.5 & 28 & 3    & 94,5 & 79,3 & $ \pm 1.3 $ & 11   & $ (3.4 \pm 2.7) \times 10^{ 2 } $ & 3; 18  \\
                               
\textit{Seg II} & 02 ~ 19 ~ 16 & +20 ~ 10 ~ 31 & $ 35 \pm 2 $  & 157.0 & -31.1 & 41 & 5    & 54,8 & 67,6 & $ \pm 2.5 $ & 5   & $ (8.6 \pm 2.7) \times 10^{ 2 } $ & 5  \\
                               
UMa & 10 ~ 34 ~ 49 & +51 ~ 55 ~ 48 & $ 100 \pm 4 $ & 162.0 & 51.3 & 105 & 34; 29; 24    & -6,9 & -0,3 & $ \pm 1.4 $ & 29   & $ (1.4 \pm 0.4) \times 10^{ 4 } $ & 18  \\
                               
UMa II & 08 ~ 51 ~ 30 & +63 ~ 08 ~ 22 & $ 30 \pm 5 $  & 159.7 & 30.4 & 37 & 35    & -29,0 & -17,1 & $ \pm 1.9 $ & 29   & $ (3.3 \pm 1.0) \times 10^{ 3 } $ & 35; 18; 22  \\
                               
Wil1 & 10 ~ 49 ~ 22 & +51 ~ 03 ~ 10 & $ 41 \pm 6 $ & 164.4 & 48.8 & 46 & 33; 8    &                 &                &  &        & $ (1.1 \pm 0.6) \times 10^{ 3 } $ & 33; 18  \\ \hline

{\bf Classical:}\\

Car$^{\dagger}$ & 06 ~ 41 ~ 37 & -50 ~ 58 ~ 00 & $ 101 \pm 5 $ & 255.2 & -21.7 & 103 & 19    & 22,5 & -4,9 & $ \pm 3 $ & 19   & $ 4.5   \times 10^{ 4 } $ & 19  \\
                               
Dra$^{\dagger}$ & 17 ~ 20 ~ 19 & +57 ~ 54 ~ 48 & $ 82 \pm 6 $ & 93.5 & 34.6 & 82 & 19    & -112,3 & -87,7 & $ \pm 2 $ & 19   & $ 2.8   \times 10^{ 5 } $ & 19  \\
                               
For$^{\dagger}$ & 02 ~ 39 ~ 59 & -34 ~ 27 ~ 00 & $ 138 \pm 8 $ & 230.0 & -63.4 & 140 & 19    & -29,2 & -40,4 & $ \pm 3 $ & 19   & $ 1.6   \times 10^{ 7 } $ & 19  \\
                               
Leo I & 10 ~ 08 ~ 27 & +12 ~ 18 ~ 30 & $ 250 \pm 30 $ & 224.7 & 48.6 & 254 & 19    & 179,9 & 165,4 & $ \pm 2 $ & 19   & $ 4.9   \times 10^{ 6 } $ & 19  \\
                               
Leo II & 11 ~ 13 ~ 29 & +22 ~ 09 ~ 12 & $ 205 \pm 12 $ & 217.5 & 66.1 & 208 & 19    & 17,0 & 9,0 & $ \pm 2 $ & 19   & $ 5.9   \times 10^{ 5 } $ & 19  \\
                               
LMC$^{\dagger}$ & 05 ~ 23 ~ 34 & -69 ~ 45 ~ 24 & $ 49 \pm 2 $  & 268.5 & -33.4 & 48 & 19    & 143,3 & 118,6 &  & 19   &$2.1\times10^9$          &37   \\
                               
SMC$^{\dagger}$ & 00 ~ 52 ~ 44 & -72 ~ 49 ~ 42 & $ 58 \pm 2 $ & 291.6 & -47.4 & 55 & 19    & 49,6 & 32,5 &  & 19      &$5.7 \times 10^8$       &37   \\
                               
Sgr$^{\dagger}$ & 18 ~ 55 ~ 03 & -30 ~ 28 ~ 42 & $ 24 \pm 2 $  & 9.4 & -22.4 & 16 & 19    & 161,1 & 164,0 & $ \pm 5 $ & 19   & $ 2.0   \times 10^{ 7 } $ & 19  \\
                               
Scu$^{\dagger}$ & 01 ~ 00 ~ 09 & -33 ~ 42 ~ 30 & $ 79 \pm 4 $  & 234.6 & -81.9 & 79 & 19    & 77,9 & 73,8 & $ \pm 3 $ & 19   & $ 2.4   \times 10^{ 6 } $ & 19  \\
                               
Sex & 10 ~ 13 ~ 03 & -01 ~ 36 ~ 54 & $ 86 \pm 4 $  & 237.8 & 40.8 & 89 & 19    & 76,9 & 56,4 & $ \pm 3 $ & 19   & $ 5.4   \times 10^{ 5 } $ & 19  \\
                               
UMi$^{\dagger}$ & 15 ~ 09 ~ 11 & +67 ~ 12 ~ 54 & $ 66 \pm 3 $  & 114.2 & 43.2 & 68 & 19    & -92,9 & -71,7 & $ \pm 2 $ & 19   & $ 3.1   \times 10^{ 5 } $ & 19  \\

\hline
\end{tabular}
\end{center}
Notes to the table: Data for the MW satellites used for fitting the
DoS.  Seg~1 and~2 (marked in \textit{italics}) are included in this
list for reference, but they have not been included in the fitting
because they appear to be diffuse star clusters \citep{Niederste09}.
The positions are given both in Heliocentric coordinates (right
ascension $\alpha_{2000}$, declination $\delta_{2000}$, and
Heliocentric distance $r_{\rm{helio}}$ for epoch J2000.0) and in
Galactocentric coordinates assuming the Sun to have a distance of 8.5
kpc from the MW centre. $l_{\rm{MW}}$\ gives the Galactic longitude
with $0^{\rm o}$ pointing from the Galactic centre to the
Sun. $b_{\rm{MW}}$ is the latitude as seen from the Galactic centre
and $r_{\rm{MW}}$ the radial distance from the centre of the MW. The
coordinates were obtained using data from the references listed in the
column labelled Ref., where more than one source is given, the
distances to the satellites were obtained by error-weighted averaging
over the available measurements. The satellite's line-of-sight
velocities with respect to the Galactic standard of rest (GSR) are
calculated assuming the Sun to move into the direction $l =
90^{\circ}$, $b = 0^{\circ}$ (in Heliocentric, Galactic coordinates)
with a velocity of either 220 km$s^{-1}$ ($v_{\rm{GSR}}^{220}$) or 250
km$^{-1}$ ($v_{\rm{GSR}}^{250}$). The measurement-uncertainties for
the radial velocities reported in the respective papers (referred to
in column Ref.) are reproduced in the column labelled $\Delta
v$. Finally, $L_{\rm{V}}$ gives the satellite luminosities in the
photometric V-band; again uncertainty-weighted averages are quoted
when more than one reference is given in column Ref.  Data marked with
$\dagger$ have measured proper motions, listed in table 1 of
\cite{Metz08}.  $^*$:~As no distance uncertainties for Pisces II are
available in the literature, the error is estimated to be 20 percent
of the distance.  {\sl References}: 1: \cite{Adenetal09}, 2:
\cite{Belokurov 2006}, 3: \cite{Belokurovetal07}, 4: \cite{Belokurov
  2008}, 5: \cite{Belokurov 2009}, 6: \cite{Belokurov10}, 7:
\cite{Coleman 2007}, 8: \cite{de Jong 2008}, 9: \cite{de Jong 2010},
10: \cite{Dall'Ora 2006}, 11: \cite{Geha 2009}, 12: \cite{Greco 2008},
13: \cite{Ibata 2006}, 14: \cite{Koch 2009}, 15: \cite{Kollmeier
  2009}, 16: \cite{Kuehn 2008}, 17: \cite{Martin 2008a}, 18:
\cite{Martin 2008b}, 19: \cite{Mateo98}, 20: \cite{Moretti 2009}, 21:
\cite{Munoz 2006}, 22: \cite{Munoz 2009}, 23: \cite{Musella 2009}, 24:
\cite{Okamoto 2008}, 25: \cite{Sakamoto 2006}, 26: \cite{Sand 2009a},
27: \cite{Sand 2009b}, 28: \cite{Siegel 2006}, 29: \cite{SimonGeha07},
30: \cite{Walsh 2007}, 31: \cite{Walsh08}, 32: \cite{Watkins09}, 33:
\cite{Willman 2005a}, 34: \cite{Willman 2005b}, 35: \cite{Zucker
  2006a}, 36: \cite{Zucker 2006b}, 37: \cite{vandenBergh99}.
\end{sidewaystable*}

\cite{Metz07} and \cite{Metz09} employed a sophisticated fitting
routine to find the DoS. Here, an intuitive plane-fitting algorithm
and a new disc-test are introduced.  The plane-fitting algorithm leads
to perfect agreement with the results obtained by Metz et al., and the
new test allows an assessment of how discy the satellite distribution
is.

\subsection{Parameters of the DoS}
\label{ssec:DoSparameters}

A simple and straightforward method is described to calculate the DoS
parameters $l_{\rm MW}$, $b_{\rm MW}$, $D_{\rm p}$, and $\Delta$,
which are, respectively, the direction of the DoS-normal vector in
Galactic longitude and latitude, the smallest distance of the DoS
plane to the Galactic centre, and the root-mean-square height (half
the thickness) of the DoS.

The positions of satellites on the sky and their radial distances
(compiled for convenience in Table~\ref{tab:satellites}) are
transformed into a Galactocentric, cartesian coordinate system
assuming the distance of the Sun to the centre of the MW to be
8.5~kpc. The $z$-coordinate points into the direction of the Galactic
North Pole and the Sun lies in the MW disk plane.

The 3D coordinates are projected into two dimensions, plotting $z$
against a projection onto a plane defined by the Galactic longitude
$l_{\rm{MW}}$.  This resembles a view of the MW satellite system as
seen from infinity and from within the MW disc plane.  The view of the
satellite system is rotated in steps of $1^\circ$. For each step, a
linear fit is made to the projected satellite distribution. The
linear fit is determined using the least squares method, demanding the
satellite-distances, as measured perpendicularly to the fitted line,
to become minimal. This line constitutes a plane seen edge-on in the
current projection.  The two free parameters of the fit are the
closest distance from the MW centre, $D_{\rm{P}}$, and the inclination
$b_{\rm{MW}}$ of the normal vector to the z-axis 
(a polar plane has $b_{\rm MW}=0^\circ$). 
The plane-normal-vector's longitude is $l_{\rm{MW}}$, given by the
projection. The fits are performed for each angle $l_{\rm{MW}}$
between $0^\circ$ and $360^\circ$. After half of a rotation, the view
is identical to the one of $180^\circ$ before, mirrored along the
z-axis.

For each angle $l_{\rm{MW}}$, the root mean square (RMS) height,
$\Delta$, of the satellite distribution around the fitted line is
determined. The normal vector to the best-fit disc solution (the DoS) to
the full 3-dimensional distribution of the MW satellites is then given
by those $l_{\rm{MW}}$ and $b_{\rm{MW}}$ that give the smallest RMS
height $\Delta_{\rm{min}}$.

To account for the uncertainties in the distance of the satellites, the major
source of error, the procedure is repeated~1000 times. Each time, the
radial position of each satellite is randomly chosen from a normal
distribution centered on the satellite's radial distance. It has a
standard deviation given by the distance uncertainties to the
satellite galaxies.  Once a realisation with varied radial distances
is set up, the coordinate transformation into the Galactic coordinate
system is performed. The parameters of the best fits are determined
for each realisation. Their final values are determined by averaging
the results of all realisations, the standard deviations of their
values are adopted as the uncertainties in the fits.  

Fitting all~24 currently known satellite galaxies within a distance of 
254~kpc from
the MW, the minimum disc height is found to be $\Delta_{\rm{min}} =
28.9 \pm 0.6~\rm{kpc}$. This is more than 14$\sigma$ away from the
maximum height of $\Delta_{\rm{max}} = 55.7 \pm 1.3~\rm{kpc}$
obtained at a $90^{\rm o}$ projection of the data. {\sl Thus, the DoS
  is highly significant.}  The position of the minimum height gives
the best-fit disc, the DoS. The normal vector defining the DoS
points to $l_{\rm{MW}} = 156^\circ.4 \pm 1^\circ.8$ and has an
inclination of $b_{\rm{MW}} = -2^\circ.2 \pm 0^\circ.6$, i.e. is
nearly perfectly polar.  $D_{\rm{P}}$, the closest distance of the DoS
from the MW centre, is~$8.2 \pm 1.0~\rm{kpc} \ll \Delta_{\rm min}$.

\begin{figure}
\hspace{-5mm}
\includegraphics[angle=0,scale=0.70]{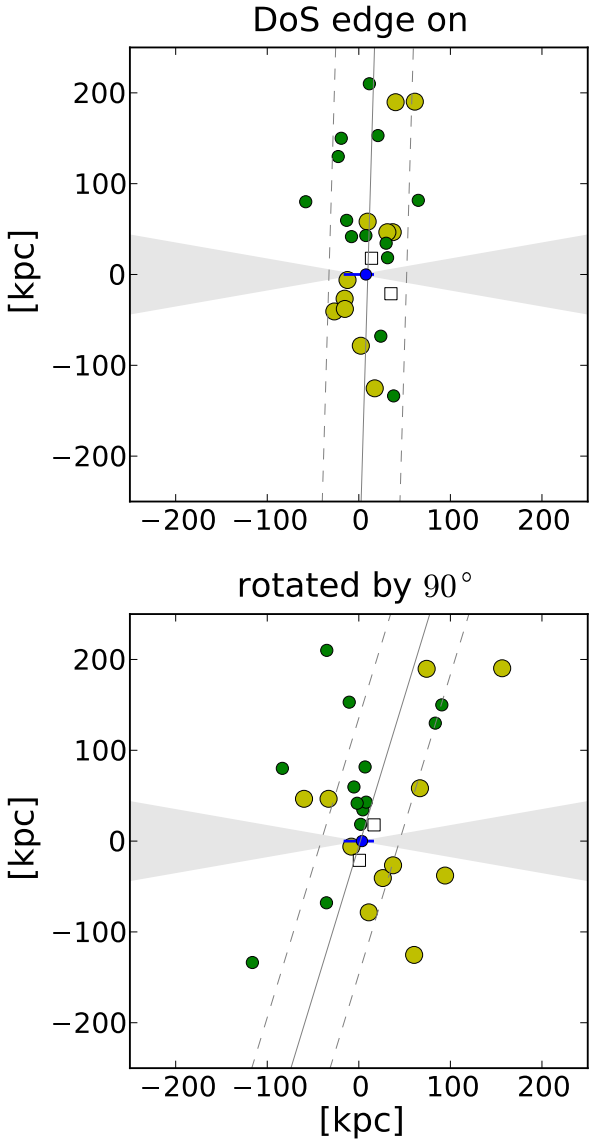}
\caption{Parameters of the MW DoS: the 3-D distribution of the MW
  satellite galaxies. The 11~classical satellites are shown as large
  (yellow) circles, the 13~new satellites are represented by the
  smaller (green) dots, whereby Pisces~I and~II are the two southern
  dots. The two open squares near the MW are Seg~1 and~2; they are not
  included in the fit because they appear to be diffuse star clusters
  nearby the MW, but they do lie well in the DoS. The
  obscuration-region of $\pm 10^{\circ}$ from the MW disc is given by
  the horizontal gray areas. In the centre, the MW disc orientation is
  shown by a short horizontal line, on which the position of the Sun
  is given as a blue dot.  The near-vertical solid line shows the best
  fit (seen edge-on) to the satellite distribution at the given
  projection, the dashed lines define the region $\pm 1.5 \times
  \Delta_{\rm{min}}$, $\Delta_{\rm min}$ being the RMS-height of
  the thinnest DoS ($\Delta_{\rm min}=28.9\;$kpc in both panels).
  {\sl Upper panel}: an edge-on view of the DoS. Only three of the 24
  satellites are outside of the dashed lines, giving $N_{\rm{in}} =
  21$, $N_{\rm{out}} = 3$ and thus a ratio of ${\cal R} = N_{\rm
    in}/N_{\rm out} = 7.0$.  Note the absence of satellites {\sl in
    large regions of the SDSS survey volume} (upper left and right
  regions of the upper panel, see also fig.~1 in \citealt{Metz09} for
  the SDSS survey regions).  {\sl Lower panel}: a view rotated by
  $90^\circ$, the DoS is seen face-on. Now, only 13 satellites are
  close to the best-fit line, 11~are outside, resulting in ${\cal R} =
  1.2$. Note that by symmetry the Southern Galactic hemisphere ought
  to contain about the same number of satellites as the Northern
  hemisphere. Thus, {\sl The Stromlo Milky Way Satellite Survey} is
  expected to find about eight additional satellites in the Southern
  hemisphere.
\label{fig:discfit}}
\end{figure}

\subsection{A novel disc test}

Another test to determine whether the satellite galaxies are distributed
in a disc can be performed by comparing the number of satellites near
the plane to the number further away: Let $N_{\rm{in}}$ be the number
of all satellites that have a perpendicular distance of less than~1.5
times the minimal disc height $\Delta_{\rm{min}}$ from the
line-fit. Accordingly, $N_{\rm{out}}$ represents all satellites further
away. Both $N_{\rm in}$ and $N_{\rm out}$ are determined for each rotation
angle, measuring distances from the line (i.e. plane viewed edge-on in
the given projection) that fits the distribution in the given
projection best. This is illustrated in Fig.~\ref{fig:discfit}. It
shows an edge-on view of the best-fit plane, along with a view
rotated by $90^{\rm o}$.  Both views see the disc of the MW edge-on.

\begin{figure}
% \hspace{-5mm}
\includegraphics[angle=0,scale=0.55]{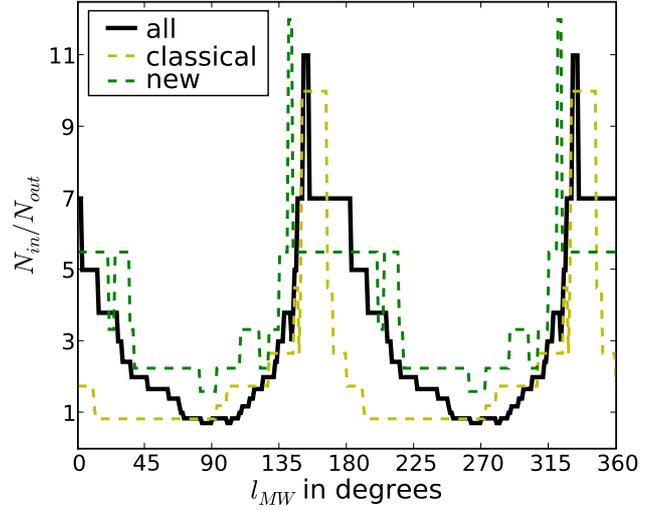}
\caption{Testing for the existence of the DoS. The behaviour of ${\cal
    R}$ for each view of the MW, given by the Galactic longitude of
  the normal vector for each plane-fit.  ${\cal R}=N_{\rm in}/N_{\rm
    out}$ is the ratio of the number of satellites within $1.5 \times
  \Delta_{\rm{min}}$ ($\Delta_{\rm min}=28.9\;$kpc), $N_{\rm{in}}$, to
  those further away from the best-fit line, $N_{\rm{out}}$,
  calculated for all~24 known satellites, as well as for the fits to
  the 11~classical and the 13~new satellites separately (taking their
  respective RMS heights as the relevant $\Delta_{\rm min}$). The
  disc-like distribution can be clearly seen as a strong peak close to
  $l_{\rm{MW}} = 150^\circ$.  Note that the position of the peaks are
  close to each other for both subsamples separately. This shows that
  the new satellite galaxies independently define the same DoS as the
  classical satellite galaxies.
\label{fig:rfit}}
\end{figure}

Figure~\ref{fig:rfit} shows the ratio of galaxies found within the DoS
to those outside (solid black line), ${\cal R} = N_{\rm{in}} /
N_{\rm{out}}$. The situation is shown for the unvaried distances.  If
the MW satellites were distributed in a disc, ${\cal R}$ would
approach a maximum when looking edge-on, while it will rapidly
decrease once the projection is rotated out of the disc plane. It is a
good test to discriminate a disc-like distribution from a spheroidal
one. The latter would not lead to much variation in the ratio.

It can be seen that ${\cal R}$ approaches a maximum close to the
best-fit $l_{\rm{MW}}$.  At the maximum, only two of the 24 satellite
galaxies are found outside of the required distance from the disc.
The maximum ${\cal R}$ is thus 11.0, situated only a few degrees away
from the $l_{\rm{MW}}$ that gives the smallest height.  This has
to be compared to the broad minimum of ${\cal R}\approx 1$. The
disc-signature is obvious, proving the existence of a DoS that
incorporates the new satellites found in the SDSS.

\subsection{Classical versus new satellites: is there a DoS in both cases?}
\label{ssec:newDoS}

In addition to the above analysis of all~24 known MW satellites, the
analysis is also carried out separately for two distinct subsamples:
the~11 classical, most-luminous satellite galaxies and the 13~new
satellites discovered mostly in the SDSS. Each of them uses an own
minimal height, given by the subsample distribution, in determining
${\cal R}$. If all satellite galaxies follow the same distribution,
given by the DoS, a separate fitting should lead to similar
parameters. If, on the other hand, the new (mostly ultra-faint)
satellites follow a different distribution, then this would challenge
the existence of a DoS. {\sl It is worth emphasising that while the
  brightest satellites in a $\Lambda$CDM model of a MW-type halo may
  exceptionally form a weak disc-like structure
  \citep{Libeskindetal09}, none of the existing CCM-based
    theoretical satellite distributions predict the whole luminous satellite
  population to be disc-like.}

Furthermore, comparing the results for the classical~11 satellites
with the ones obtained by the more sophisticated fitting technique
used by \cite{Metz07} is a good test to check whether the present
technique gives reliable results.

The graphs for both subsamples are included in Fig.~\ref{fig:rfit},
the results for classical satellites are represented by the dashed
yellow, the new (SDSS) satellite galaxies by the dashed green
line. Both are in good agreement not only with the combined sample,
but also with each other. They peak at their best-fit $l_{\rm{MW}}$,
with each of them having an $N_{\rm{out}}$ of only one galaxy at the
peak.

Applying the technique presented in Sect.~\ref{ssec:DoSparameters} to
calculate the DoS parameters, the new satellites have a best-fit disc
with a normal vector pointing to $l_{\rm{MW}} = 151^\circ.4 \pm
2^\circ.0$, only five degrees away from the direction that was
obtained by considering all known MW satellites.  The inclination is
$b_{\rm{MW}} = 9^\circ.1 \pm 1^\circ.0$, again an almost perpendicular
orientation of the DoS relative to the MW disc, being only 11~degrees
away from the value determined before.  The derived RMS height is
$\Delta_{\rm{min}} = 28.6 \pm 0.5~\rm{kpc}$, essentially identical to
the one given by all satellite galaxies. The minimum distance from the
MW centre is $D_{\rm{P}} = 18.3 \pm 1.3~\rm{kpc}$.

The fitting to the~11 classical satellites leads to results that are
in very good agreement, too.  The best-fit position for the 11
classical satellites is $l_{\rm{MW}} = 157^\circ.6 \pm 1^\circ.1$ and
$b_{\rm{MW}} = -12^\circ.0 \pm 0^\circ.5$, the height is found to
be $\Delta = 18.3 \pm 0.6~\rm{kpc}$, and the closest distance to the
MW centre is $D_{\rm{P}} = 8.4 \pm 0.6~\rm{kpc}$.  This is in
excellent agreement with the results of \cite{Metz07}. In that
paper, the authors reported that $l_{\rm{MW}} = 157^\circ.3$, $b_{\rm{MW}}
= -12^\circ.7$, $\Delta_{\rm{min}} = 18.5~\rm{kpc}$, and $D_{\rm{P}} =
8.3~\rm{kpc}$.  This illustrates that the results are extremely
accurate despite employing a more simple disc-finding technique.

The agreement of the fit parameters for the two subsamples {\sl
  separately} is impressive. Two populations of MW satellite galaxies
(classical versus ultra-faint) with different discovery histories and
methods define the same DoS.  This shows that the new, faint
satellites fall close to the known, classical, DoS ($\equiv$DoS$_{\rm
  cl}$).  Even without considering the classical satellite galaxies,
the new satellites define a disc, DoS$_{\rm new}$, that has
essentially the same parameters. This confirms the existence of a
common DoS$\approx$DoS$_{\rm new}\approx$DoS$_{\rm cl}$.

\subsection{The DoS -- Discussion}
\label{ssec:DoSDisc}

A pronounced DoS is therefore a physical feature of the MW system. But
what is its origin? Is the existence of both the classical-satellite
DoS$_{\rm cl}$ and the new-satellite DoS$_{\rm new}$, such that
DoS$_{\rm new}\approx\;$DoS$_{\rm cl}$, consistent with the CCM?

It has been suggested that the highly anisotropic spatial satellite
distribution maps a highly prolate DM halo of the MW that would need
to have its principal axis oriented nearly perpendicularly to the MW
disc \citep{Hartwick00}. However, there is still much uncertainty and
disagreement as to the shape and orientation of the MW DM halo:
\cite{Fellhaueretal06} used the bifurcation of the Sagittarius stream
to constrain the shape of the MW DM halo to within about 60~kpc, finding
it to be close to spherical.  The measurement of the shape of
the DM halo of the MW within 60~kpc by \cite{LMJ09}, also based on the
Sagittarius stream, suggests that the DM halo is triaxial, but with
major and minor axes lying within the plane of the MW disc.  The DM
halo of the MW would therefore not trace a similar three-dimensional
structure as the satellites, unless the major axis of the MW halo
changes its orientation by about 90~degrees beyond~60~kpc and becomes
essentially disc-like (i.e. highly oblate).  \cite{LM10} find a new
slightly oblate solution to the MW DM halo shape at distances from
20~to~60~kpc. In this solution, the minor axis points along the line
Sun--MW-centre suggesting a similar orientation of this extra
potential as the DoS. The authors emphasise that this model is not
strongly motivated within the current CDM paradigm, it merely serving
as a ``numerical crutch''. Given this disagreement about the shape and
orientation of the MW DM halo, a significant future observational and
theoretical effort to clarify the situation is needed.

An additional issue for the CCM is that the normal to the DoS is
defined mostly by the outermost satellites, while the direction of the
average orbital angular momentum vector is defined by the innermost
satellites for which proper motions have been measured. Both, the
normal and the average orbital angular momentum vector are nearly
co-aligned implying a strong degree of {\sl phase-space correlation}
between the satellites such that the DoS is rotating \citep{Metz08}.
This rotational DoS is not expected if the satellites merely trace
the MW DM halo, because they would have independent infall histories
and would therefore not be correlated in phase space.

This phase-space feature has been addressed by \cite{Libeskindetal09}.
In a thorough analysis of structure formation on MW-galaxy scales,
they show that the MW constitutes an improbable but possible
constellation of CDM-dominated satellites about a MW-type disk galaxy,
the satellites having (of course) independent infall and accretion
histories.

They analyse an N-body sample of~30946 MW-mass DM host halos with mass
in the range $2\times 10^{11}\,M_\odot$ to $2\times 10^{12}\,M_\odot$
for the properties of their substructure distribution. They first
select from this sample only those halos that host a galaxy of similar
luminosity as the MW (specifically, galaxies more luminous in the
V-band than $M_V=-20$).  From this remaining sample of 3201 (10~per
cent) hosts, they select those that contain at least~11 luminous
satellites, leaving~436 (1.4~per cent) host halos. In this sample of
436 systems, about~30~per cent have~6~luminous satellites with orbital
angular momenta aligned to a degree similar to that of the MW system. Thus,
only 0.4~per cent of all existing MW-mass CDM halos would host a
MW-type galaxy with the right satellite spatial distribution. As the
authors point out, this probability of $4\times 10^{-3}$ that the DM
model accounts for the observed MW-type satellite system would be
lower still if proper motion measurements of additional satellites
affirm the orbital angular momentum correlation highlighted by
\cite{Metz08}, or if the satellites that may be discovered in the
southern hemisphere by the {\sl Stromlo Milky Way Satellite Survey}
\citep{Jerjen10}\label{note1}\footnote{http://www.mso.anu.edu.au/$\sim$jerjen/SMS\_Survey.html}
also lie within the DoS. All 13~new satellites define the
same DoS as the~11 classical ones, and furthermore, the latest
additions in the southern Galactic hemisphere also lie in the DoS
(Sect.~\ref{ssec:newDoS}), {\sl suggesting that the DM hypothesis is much
less likely than~0.4~per cent to be able to account for the MW
satellite system in MW-type DM halos}.

\cite{Li08} and \cite{DOnghia08} propose an interesting alternative
solution to the {\sl satellite phase-space correlation problem}: they
suggest that the correlation is caused by the infall of groups of
DM-dominated dwarf galaxies.  Unfortunately, this proposition is
challenged by all known nearby groups of dwarf galaxies being
spatially far too extended to account for the thinness of the DoS
\citep{Metz09b}. It may be thought that the groups that have fallen in
correspond to compact dwarf groups that do not exist any longer
because they have subsequently merged.  But this is compromised by the
observation that their putative merged counterparts in the field do
not seem to exist \citep{Metz09b}. Indeed, \cite{Klimentowskietal09}
model a MW-type system and deduce ``... that such a disc is probably
not an effect of a group infall unless it happened very recently''
(their section~4.2.2). Furthermore, this notion would seem to imply
that dwarf galaxy groups are full of dSph galaxies, while the pristine
(before group infall) MW halo would have formed none, in conflict with
the observed morphology-density relation (e.g. \citealt{OT00}).

It needs to be emphasised that the DM-based models have so far not
addressed the issue that the DoS lies nearly perpendicularly to the MW
disc; DM-based models need to {\sl postulate} that this occurs, and it
may indeed simply be chance. The combined probability that a DM-based
model accounts for the observed MW-type system, which has the
properties that the satellites have correlated angular momenta and
form a DoS highly inclined to the baryonic disc of the host galaxy,
cannot currently be assessed but is, by logical implication, smaller
than $4\times 10^{-3}$.

But perhaps the MW is a very special system, an outlier within the
DM-based framework?  This possibility can be assessed by considering
the nearest MW-similar DM halo.  It hosts a similar disc galaxy,
Andromeda, which has a similar satellite system to the MW but is
however richer and more complex, and has a larger bulge mass than the
MW (Fig.~\ref{fig:correl}).  Andromeda may also have a DoS (\citealt{KG06},
see also fig.~4 in \citealt{Metz09})\footnote{Note that the rich
  satellite system of M31 may have a sub-population of satellites in a
  disc-like structure \citep{Metz09}.} suggesting that these satellite
distributions may not be uncommon among MW-type DM halos.

Thus, a Local Group consisting of two dominant DM halos of similar
(MW-type) mass would have a likelihood of 0.4~per cent times 1.4~per
cent, i.e. $5.6\times 10^{-5}$, to appear with two MW-type
disc galaxies, one of them having a pronounced rotating DoS with~11 or
more luminous satellites, and the other having at least~11 luminous
satellites.

\subsection{Invariant baryonic galaxies}
\label{ssec:baryonic_invariant}

The \cite{Libeskindetal09} analysis, described in
Sect.~\ref{ssec:DoSDisc}, also shows that about 10~per cent of MW-type
DM halos would host a MW-luminous galaxy, the 90~per cent of others would
presumably host galaxies with lower luminosities suggesting a large
variation between DM halo and luminous galaxy properties.  This
however, appears to be a problem considering the properties of
observed disc galaxies. By using a principal component analysis on
hundreds of disc galaxies, \cite{Disneyetal08} demonstrate that
observed disc galaxies are simple systems defined by one underlying
parameter, rather than about six if the galaxies were immersed in DM
halos.  Citing additional similar results, \cite{vandenBergh08} and
\cite{Gavazzi09} reach the same conclusion, as well as
\cite{Gentileetal09} and \cite{Milgrom09}.  This is further supported
by an entirely independent study of star-forming galaxies, which again
shows a remarkably small variation of behaviour \citep{PAK09b}.  The
discovery that the ratio of DM mass to baryonic mass
within the DM core radius is constant for galaxies
(Sect.~\ref{ssec:nonNewt} below) is another statement of the same
effect.

The small amount of variation for disc galaxies thus appears to be very
difficult to reconcile with the large variation inherent in the DM
model, as quantified by the \cite{Libeskindetal09} analysis: 90~per
cent of MW-mass DM halos would have disc galaxies that differ
substantially in luminosity from the MW in the CCM, and yet the
closest neighbour, Andromeda, is similar to the MW. This is the {\sl
  invariant-baryonic-galaxy problem}.

{\sl Summarising Sect.~\ref{sec:DoS}}, the CCM is highly significantly
challenged by the spatial distribution of MW satellite galaxies and by
the similarity of rotationally supported galaxies. The {\sl
  phase-space correlation problem} of the classical satellites is
enhanced significantly after the inclusion of the new ultra-faint
satellites, and the Local Group enhances the {\sl invariant
    baryonic galaxy problem}.

\section{The origin of dSph and dE galaxies: The Fritz Zwicky Paradox, 
an alternative proposition and deeper implications}
\label{sec:tdgs}

What has been learned so far? The DM-mass--luminosity data of MW dSph
satellite galaxies appear to be in conflict with the CCM results, and the
mass function of DM masses of the dSph satellites is not in good
agreement with the mass function of luminous sub-halos calculated
within the CCM.  The correlation bulge-mass versus satellite-number is
tentative (for having only three points) but will very likely pass the
test of time because the error bars allow for a conclusive
significance test.  The two quantities appear to be physically related
as indicated strongly by the Local Group data and also extragalacitc
surveys, but clearly much more work needs to be done both
observationally and theoretically to refine the implied correlation.
The highly pronounced phase-space correlation of the MW satellites
means that any formation mechanism must have involved correlated
orbital angular momenta of the satellites.

Given that the formation of a bulge involves highly dissipative
processes, it emerges that a highly dissipative process seems to have
formed the bulge and the correlated orbital angular momenta of the
satellites. This leads to the real possibility that the origin of both
the MW bulge and its satellite population is related to a
galaxy--galaxy encounter. Indeed, it is well known and documented that
galaxy encounters lead to the formation of bulges {\sl and} tidal arms
that can host the formation of tidal-dwarf galaxies (TDGs). These are
then naturally correlated in phase space. Since the bulge and the
satellites of the MW are about 11~Gyr old, we are led to the scenario
that the proto-Galaxy may have had a major encounter or merger about
11~Gyr ago during which the bulge and the satellites formed
\citep{Pawlowskietal2010}.  \cite{Wetzsteinetal07} demonstrated in a
series of numerical models that the number of TDGs increases indeed
with the gas fraction of the pre-collision galaxy. This is relevant to
galaxy encounters at a high redshift, where galaxy encounters are
expected to have been frequent.

Noteworthy is that a scenario for the origin of dSph satellite
galaxies along the above lines had been suggested already before the
DM hypothesis was widely accepted, namely that they may be ancient
TDGs \citep{LyndenBell76,LyndenBell83,Kunkel79}.  This proposition can
naturally account for their correlated phase-space distribution in the
form of a rotating disc-like distribution (Sect.~\ref{sec:DoS}), and
would lend a new viewpoint on the difficulty of understanding the
properties of the MW dSph satellites as DM sub-halos documented
above. Indeed, in a famous conjecture, Fritz Zwicky (\citealt{Zw56},
on p. 369) states that new dwarf galaxies form when galaxies
interact. As shown here this leads to a contradiction with
observational data when this conjecture is combined with his other
famous conjecture according to which the masses of galaxies are
dominated by Dark Matter \citep{Zw56}. This contradiction is referred
to as the Fritz Zwicky Paradox.

\subsection{The evolution of TDGs}
\label{ssec:tdgsevol}

A natural way to explain the satellite phase-space correlation as well
as the bulge-satellite relation is thus to identify the dSph satellite
galaxies of the MW with a population of ancient TDGs that probably
formed during a gas-rich encounter between the early MW and another
galaxy.  But if they all formed at the same time, how can the
different chemical properties and star-formation histories of the
different dwarf galaxies then be explained within this scenario? If the DM
hypothesis is not viable for the MW satellite population, how can the
high mass-to-light ratios of the satellites be explained?

It is known that the satellite galaxies all have ancient populations
of an indistinguishable age \citep{Grebel08}, perhaps being created
when the TDGs were born. Or, the ancient population comes from the
precursor host galaxy.  TDGs may also form with globular clusters as
long as the star-formation rate surpasses a few~$M_\odot$/yr
for~10~Myr \citep{WKL04}.  The chemo-dynamical modelling by
\cite{Recchietal07} has shown that once a TDG (without DM) forms it is
not natural for it to blow out the gas rapidly. Rather, the
rotationally-supported small gas-rich discs of young TDGs begin to
evolve through self-regulated star formation either until their gas is
consumed or removed through ram-pressure stripping.  Consequently,
their internal evolution through star formation can be slow and individual,
such that TDGs that formed during one encounter event can exhibit
different chemical properties many~Gyr after their formation.
Removal of the interstellar medium from the TDG through ram-pressure
takes about half to a few orbital times, which is typically one to a
few Gyr after formation. This time scale is consistent with the
observed cessation of star formation in most MW~dSph satellites
\citep{Grebel99}. The TDGs that have remained at large distances from
their hosts retain their gas and appear as dIrr galaxies
\citep{Hunteretal00}.  Once formed, TDGs cannot fall back onto their
hosts and merge since dynamical friction is insignificant for them. A
TDG may be dispersed (but not accreted) if it happens to be on a near
radial orbit, which, however, is unlikely given the torques acting on
the tidally expelled material from which the TDG forms during the
encounter.

If the dSph satellites are ancient TDGs then understanding their
internal kinematics remains a challenge though because TDGs do not
contain significant amounts of DM \citep{BarnesHernquist92,
  Wetzsteinetal07, Bournaud07b, Gentile07, Milgrom07}.  However, the
inferred large $M/L$ ratios of dSph satellites (and especially of the
ultra-faints) may not be physical values but may be a
misinterpretation of the stellar phase-space distribution within the
satellite. If this were to be the case then the absence of a
``DM-mass"-luminosity relation (Sect.~\ref{sec:ML}) for dSph
satellites would be naturally understood.

The following gedanken-experiment illustrates that this could be the
case.  An unbound population of stars on similar orbits, each
  slightly inclined relative to the other orbits, will reconfigure at
 apogalacticon and an observer would see a stellar phase-space
density enhancement and would also observe a velocity dispersion. The
$M/L$ ratio calculated from the observed velocity dispersion would not
be a true physical $M/L$ ratio.  Models related to this idea have been
studied by \cite{Kuhn93}. Moreover, resonant orbital coupling can
periodically inflate kinematically measured $M/L$ values
\citep{KM89,KSH96}.  Fully self-consistent Newtonian N-body models
have demonstrated that unphysically high $M/L$ ratios arise indeed if
TDGs are allowed to orbit a host galaxy sufficiently long such that
the remaining stellar population within the ancient TDG adopts a
highly non-isotropic phase-space distribution function
\citep{Kroupa97,KlKr98,MK07}. These models suggest that it may be
wrong to use an observed velocity dispersion to calculate a mass for a
dSph satellite. Thus, tidal shaping of TDGs over a Hubble time can
produce remnant objects that have internal highly-anisotropic stellar
phase-space distributions that would be falsely interpreted by an
observer as corresponding to a high $M/L$ ratio, as explicitly shown
by \cite{Kroupa97}.  Intriguingly, these models reproduce the gross
physical parameters of dSph satellites well \citep{MK07}, and thus
constitute the simplest available stellar dynamical solutions of dSph
satellites constructed without fine-tuning.

\begin{figure}
\hspace{-5mm}
\includegraphics[angle=0.5,scale=0.499]{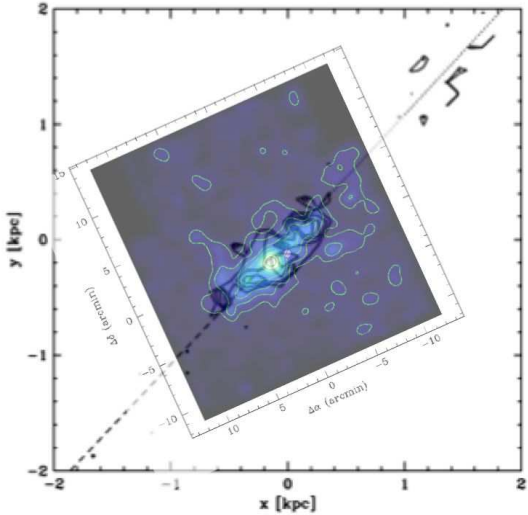}
\caption{Model RS1-5 from \cite{Kroupa97} (on the kpc grid) is plotted
  over the surface brightness contours of Hercules by
  \cite{Coleman2007} (celestial coordinate grid). The dashed and
  dotted curve are, respectively, the past and future orbit of RS1-5.
\label{fig:hercules}}
\end{figure}

It is indeed remarkable how model RS1-5 of \cite{Kroupa97}, shown here
as a snapshot (Fig.~\ref{fig:hercules}), is an essentially perfect
match to the dSph satellite Hercules (see fig.~2 in \citealt{Coleman2007})
discovered~10 years later by \cite{Belokurovetal07}.  The half-light
radius is 180~pc in the model and 168~pc for Hercules, RS1-5 has a
velocity dispersion of about 2.8~km\,s$^{-1}$ (table~2 in
\citealt{Kroupa97}), while Hercules has a measured velocity dispersion
of $3.72\pm0.91$~km\,s$^{-1}$ \citep{Adenetal09}, and the inferred
mass-to-light ratio that one would deduce from velocity dispersion
measurements based on the assumption of equilibrium is about $200$ in
both cases.  Both RS1-5 and Hercules have luminosities agreeing within one
order of magnitude (the model being the brighter one), yet RS1-5 has no DM.

The TDG models for dSph satellites proposed by
\cite{LyndenBell76,LyndenBell83} and \cite{Kunkel79} and calculated by
\cite{Kroupa97} and \cite{KlKr98}, which are based on observed
properties of TDGs, thus lead to a population of ancient TDGs that are
in reasonable agreement with the observed luminosities, dimensions,
and $M/L$ ratios of dSph satellites \citep{MK07}.  These model-dSph
satellites require no fine tuning of parameters but only assume the
formation about~10~Gyr ago of about $10^7\,M_\odot$ heavy TDGs
composed purely of baryonic matter. This theoretical framework of
satellite galaxies does not imply any relation between luminosity and
(wrongly inferred) ``dynamical mass", in agreement with the lack of
this relation (Sect.~\ref{sec:ML}). And it would naturally explain why
the mass function of luminous DM sub-halos cannot account for the
observations (Sect.~\ref{sec:mfn}).  Within Newtonian dynamics, this
dynamical modelling over many orbits around the MW DM halo has
demonstrated that even low-mass satellites do not easily disrupt
unless they are on virtually radial orbits \citep{Kroupa97, MK07}.

{\sl Summarising Subsect.~\ref{ssec:tdgsevol}}, the physics of TDG
formation and evolution is sufficiently well understood to conclude
that 1) {\sl once formed at a sufficient distance from the host, TDGs
  will take an extremely long time to dissolve, if at all}; 
and 2) the TDGs formed will naturally lead to a population of ancient TDGs
that resemble dSph satellites.  A bulge-mass--number of satellite
correlation and a DoS arise naturally in this scenario.

\subsection{On the substructure problem}
\label{ssec:sub}

The MW dSph satellites can therefore be understood as ancient TDGs
that formed within a DM universe. But on the other hand, the extensive
modelling within the CCM strictly implies, if DM is cold or warm (but
not hot), that MW-luminous galaxies must be accompanied by hundreds
(with a slight dependence on the cold or warm nature of DM) of shining
albeit faint satellites, which are not of tidal origin
\citep{Knebeetal08,Maccioetal09,Busha09,Koposovetal09}.  For example,
\cite{Tollerudetal2008} conjecture that ``there should be between ~300
and~600 satellites within $D=400$~kpc of the Sun that are brighter
than the faintest known dwarf galaxies and that there may be as many
as~1000, depending on assumptions.''  Deep follow-up observations of
the low S/N ultra-low-luminosity satellite candidates introduced by
\cite{Walshetal09} show that these are not dSphs as a population.
These results show that there is not a significant number of missing,
ultra-low-luminosity satellites ($M_V > -2, D < 40\,$kpc) in the SDSS
footprint, i.e. an area covering half of the Northern hemisphere
(Jerjen et al., in prep.).  This may be a problem because of the
$\Lambda$CDM prediction that there should be a dozen additional
satellites ($M_V<0, D<40$~kpc) in a quarter celestial sphere
(e.g. fig.~4 in \citealt{Koposovetal09}; see also \citealt{Cooper10}).

If the dSph satellites are ancient TDGs stemming from an early
gas-rich encounter involving the proto-MW and probably contributing a
collision product to the MW bulge (see Sect.~\ref{sec:origin}), then
this would mean that the MW would have a severe substructure problem
as there would not be any satellites with DM halos less massive than
about $10^{10}\,M_\odot$ with stars, in conflict with DM predictions
provided by, e.g., \cite{Knebeetal08}, \cite{Diemand08}, \cite{Busha09},
 \cite{Maccioetal09}, and \cite{Koposovetal09}. 
Perhaps a few dSph satellites are
ancient TDGs, such as the classical or nine brightest satellites, and
the remainder are the DM dominated sub-halos?  This possibility is
unlikely, because the new satellites span the same DoS
(Sect.~\ref{ssec:newDoS}) and because they do not form a population
with physical properties that differ distinctly from those of the
classical satellites (e.g. \citealt{Strigari08}).

{\sl Summarising Subsect.~\ref{ssec:sub}}, based purely on the
existence of the satellite phase-space correlation and the formation
and survival of TDGs in a hierarchical structure formation framework
the Fritz Zwicky Paradox emerges and the validity of the DM hypothesis
must be questioned, because the dSph satellites cannot be two types of
object at the same time, namely DM-dominated sub-structures and
ancient DM-free TDGs.

\subsection{Early-type galaxies}
\label{ssec:dE}
But if TDGs account for the dSph satellites of the MW, would they then
not also be an important population in other environments?  The
production of TDGs in the CCM has been calculated by
\cite{OT00}. Intriguingly, they find that TDGs naturally match the
observed number of dE galaxies in various environments. The result of
\cite{OT00} is rather striking, since they find that within the CCM
framework only one to two long-lived (i.e., bright) TDGs need to be
produced on average per gas-dissipational encounter to cater for the
population of dwarf elliptical (dE) galaxies and for the
density--morphology relation in the field, in galaxy groups and in
clusters\footnote{Note that \cite{OT00} write: ``Adopting the galaxy
  interaction scenario proposed by Silk \& Norman, we find that if
  only a few dwarf galaxies are formed in each galaxy collision, we
  are able to explain the observed morphology-density relations for
  both dwarf and giant galaxies in the field, groups of galaxies, and
  clusters of galaxies.''  They also state ``The formation rate of
  TDGs is estimated to be~$\sim1-2$ in each galaxy interaction.'' and
  proceed to compare this number with the actually observed number of
  TDGs born in galaxy encounters. This statement is at odds with the
  quotation in \cite{Bournaud09}.}.

Viewing dE galaxies as old TDGs would be consistent with them
deviating from the mass-radius, $M(r)$, relation of pressure-supported
(early-type) stellar systems. The dE and dSph galaxies follow a
$r\propto M^{1/3}$ sequence reminiscent of tidal-field-dominated
formation.  {\sl All} other pressure-supported galactic systems
(elliptical galaxies, bulges, and ultra-compact dwarf galaxies) with
stellar mass $M>10^6\,M_\odot$ follow instead the relation $r\propto
M^{0.60\pm0.01}$ (see fig.~2 in \citealt{DHK08}, see also fig.~7 in
\citealt{Forbesetal08} and fig.~11 in \citealt{GrahamWorley08}), which
may result from opacity-limited monolithic collapse \citep{Murray09}.
Viewing dE galaxies as TDGs would also be consistent with the
observation that they have essentially stellar mass-to-light ratios
similar to globular clusters
\citep{Benderetal92,Gehaetal03,DHK08,Forbesetal08}.  If dE (baryonic
mass $>10^8\,M_\odot$) and dSph (baryonic mass $<10^8\,M_\odot$)
galaxies are old TDGs, why do they appear as different objects? That
the dE and dSph galaxies differ in terms of their baryonic-matter
density may be a result of the finding that below~$10^8\,M_\odot$
spheroidal objects on the $r\propto M^{1/3}$ relation cannot hold
their warm gas and consequently they must expand \citep{PAK09},
becoming more susceptible to tides from their host.

dE galaxies are pressure-supported stellar systems, while young TDGs
are rotationally supported \citep{Bournaudetal08}.  With a mass of
less than typically $10^9\,M_\odot$, the velocity dispersion of their
stellar populations becomes comparable to their rotational velocity
(of the order of $30$~km\,s$^{-1}$).  That a sizeable fraction of dE
galaxies show rotation, some even with spiral structure
\citep{Jerjen00, Barazzaetal02,Gehaetal03,Ferrarese06,Chilingarian09,
  Beasleyetal09}, is thus also consistent with their origin as
TDGs. For an excellent review on dE galaxies the reader is referred to
\cite{Lisker09}.

One is thus led to the following seemingly logical impasse, i.e. to
the Fritz Zwicky Paradox. In the CCM, TDGs are formed and their number
and distribution is calculated to match the number and distribution of
observed dE galaxies in the different environments. Within the CCM,
the observed luminous dwarf sub-structures are thus naturally
accounted for by TDGs. But the dE galaxies cannot be both, DM
sub-halos {\sl and} TDGs at the same time.

{\sl Summarising Subsect.~\ref{ssec:dE}}, the physical processes at play
during structure formation in the CCM imply that dE galaxies ought to
be identified as ancient TDGs.  Thus, there would be no room for
shining DM substructures.

\subsection{Deeper implications: gravitational dynamics}
\label{sec:gravdyn}

In Sects.~\ref{ssec:sub} and~\ref{ssec:dE} it has been shown that the
DM hypothesis leads to the Fritz Zwicky Pradox when accounting for the
number of satellite and dE galaxies because the formation of TDGs is
an intrinsic outcome of structure formation. In Sects.~\ref{sec:ML} to
\ref{sec:DoS} it has also been shown that the CCM seems to have a
rather major problem accounting for the observed Galactic satellites
and their internal properties. This situation suggests that
alternative ideas should be considered to help us understand the
origin of these problems, and indeed repeat the steps that had led to
a full-fledged DM framework of structure formation but with a
different outlook. Since structure formation in the DM framework
relies on Newtonian gravitation in the weak-field limit, one is
naturally led to relax insistence on Newtonian dynamics in the
weak-field limit and to consider modified gravitation theories, which
remain compatible with General Relativity in the strong field regime
and with the observed large-scale structure. We note that adopting
non-Newtonian dynamics in the weak-field limit would {\sl not}
necessarily rule out the existence of DM: on the scale of galaxy
clusters DM might still be needed, but instead of being warm or cold,
it would be {\sl hot} \citep{Angusetal09}.

\subsubsection{Non-Newtonian weak-field gravity}
\label{ssec:nonNewt}
Alternatives to Newtonian dynamics in the weak-field limit have 
been studied in great detail.  The increasingly popular
mo\-di\-fied-\-New\-tonian-\-dynamics (MOND) approach rests on a
modification of the Newtonian acceleration in the weak-field limit,
i.e. when the Newtonian acceleration $a$ is smaller than a threshold
$a_0$ \citep{Milgrom83,BekensteinMilgrom84, SM02,Bekenstein04, FB05,
  Fameyetal07, Sanders07,Sanders08, McGaugh08,Nipetal08, TC08,
  Brunetonetal09}. A modified-gravity (MOG) adding a Yukawa-like force
in the weak-field limit has also been under investigation
(\citealt{MT09,MT09b}, and references therein).  In addition, an extension of
the General Theory of Relativity to a class of alternative theories of
gravity without DM and based on generic functions $f(R)$ of the Ricci
scalar curvature $R$ have been developed and successfully applied
to the problem of galactic rotation curves (e.g. \citealt{Cap09}). For
a brief review of MOND and MOG and Milgrom's proposition on the
possible physical origin for the existence of $a_0$, the reader is
directed to the Appendix.

Both the MOND and MOG approaches have been applied to the satellite
galaxy problem with appreciable success
\citep{Milgrom95,BM00,Angus08,MT08,Hernandezetal10,McGaughWolf10}. It
has already been conclusively demonstrated that spiral galaxy rotation
curves are well recovered in MOND purely by the baryon distribution
without any parameter adjustments
\citep{SM02,McGaugh04,McGaugh05,Sanders07b}, and MOG is reported to
also do well on this account \citep{BM06}.  In contrast, the DM
approach can only poorly reproduce the vast variety of rotation
curves, and cannot explain the amazing regularities found in them
\citep{McGaugh04,McGaughetal07,Gentileetal09,Milgrom09}.  Notably, the
realisation \citep{Gentileetal09, Milgrom09} that the ratio of DM mass
to baryonic mass within the DM core radius is constant despite the
large variation in the DM--to--baryonic-matter ratio globally within
galaxies cannot be understood within the DM hypothesis.  A constant
ratio within that radius implies that the distribution of baryonic
matter is indistinguishable from that of the supposedly present DM (as
already found by \citealt{Bosma81}). This implies a hitherto not
predicted near-exact coupling between DM and baryonic matter that does
not arise naturally in the CCM, while outside that radius the effects
of DM should become noticeable \cite{McGaugh10}. The only way to
physically couple DM and baryons with each other to this degree would
be by postulating the existence of an unknown dark force that acts
only between DM particles and baryons. The modified DM cosmology would
then comprise inflation, dark matter, a dark force, and dark energy.

In MOND~models, this behaviour of gravity comes naturally.  That the
rotation curves would be purely defined by the baryonic matter
distribution in non-DM models indeed would naturally explain the later
finding based on a large sample of galaxies by \cite{Disneyetal08},
\cite{Gentileetal09}, and \cite{Milgrom09} that disc galaxies appear
to be governed by a single parameter.  Furthermore, the high
  galaxy-cluster--galaxy-cluster velocities required to obtain the
  features of the Bullet cluster have been shown to be extremely
  unlikely in the CCM (Sect.~\ref{sec:introd}), but these velocities
  are found to naturally occur in MOND \citep{AM08}. Last but not
least, the {\sl time-delay problem} of the CCM mentioned in
Sect.~\ref{sec:introd} would disappear naturally.

\subsubsection{A consistency check}
\label{ssec:nonnewton}

If it were true that the physical Universe is non-Newtonian in the
weak-field limit, then a simple test would provide a consistency check:
high dynamical mass-to-light ratios, $(M/L)_{\rm dyn}$, (derived
assuming Newtonian dynamics) would not be due to DM but due to the
dynamics being non-Newtonian in the weak-field limit and/or be due to
objects being unbound non-equilibrium systems
(Sect.~\ref{ssec:tdgsevol}).  Thus, taking MOND to be a proxy for
non-Newtonian dynamics in the weak-field limit (MOND is, arguably, the
currently available simplest alternative to Newtonian dynamics in the
weak-field limit), all systems with non-stellar $(M/L)_{\rm dyn}$
values (as derived in Newtonian gravity) would have to have internal
accelerations roughly below the MONDian value\footnote{Note that this
  statement is approximately true for all non-Newtonian gravitational
  theories since they must account for the same non-Newtonian
  phenomena in the weak-field limit.} $a_o= 3.9$~pc$/$Myr$^2$.  That
is, all pressure-supported (spheroidal) stellar systems that appear to
be dominated dynamically by DM would need to have an internal
acceleration $a<a_o$.  Note that the emphasis here is on
pressure-supported systems since rotationally supported systems have
been extensively and successfully studied in non-Newtonian
gravitational theories and because dSph and dE galaxies are mostly
pressure-supported objects.

\begin{figure}
\vspace{3mm} \includegraphics[angle=0,scale=0.48]{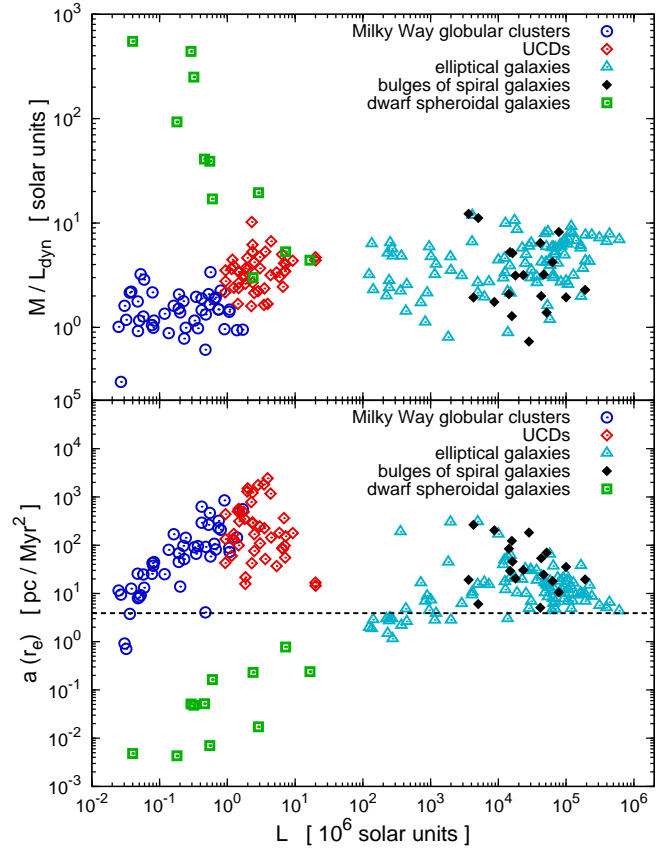}
\vspace{2mm}
\caption{{\sl Upper panel}: The dynamical $(M/L)_{\rm dyn}$ ratio
  (calculated assuming Newtonian dynamics to be valid in the
  weak-field limit) in dependence of the luminosity, $L_V$, for
  pressure-supported stellar systems following \cite{DHK08}. Note that
  here dE ($<10^{10}\,L_\odot$) and E ($>10^{10}\,L_\odot$) galaxies
  are both plotted with the same symbol.  {\sl Lower panel}: The
  Newtonian acceleration (Eq.~\ref{eq:a}) of a star located at the
  effective radius within the host system in dependence of the host
  luminosity. The dashed line is $a_0$. Note that $M/L_{\rm dyn}$ is
  high in pressure-supported stellar systems only when $a<a_0$.  In
  both panels: UCD$=$ultra compact dwarf galaxy. Comparing the upper
  and lower panels shows that evidence of DM ($M/L_{\rm dyn}>10$)
  appears only when $a<a_0$.
\label{fig:accel}}
\end{figure}

Figure~\ref{fig:accel} shows the acceleration,
\begin{equation}
a(r_e) = G\,{M \over r_{\rm e}^2} = G \frac{0.5 \Upsilon \cdot L_V}{r_e^2},
\label{eq:a}
\end{equation}
that a star inside a pressure-supported system experiences at the
effective radius, $r_e$, of its host system with luminosity spanning
$10^4$ to $10^{12}\,L_\odot$. Here $M=0.5\,\Upsilon \, L_V$ is the
stellar mass within $r_e$ and $L_V$ is the absolute V-band luminosity
in solar units.  The stellar mass-to-light ratio in the V-band is
$\Upsilon\approx3$ for collisionless systems (two-body relaxation time
longer than a Hubble time), while $\Upsilon\approx1.5$ for collisional
systems, i.e. for systems that have evaporated a significant fraction
of their faint low-mass stars by means of energy equipartition
\citep{KL08,KM09}.  Values of $(M/L)_{\rm dyn}$ as high as~10 can be
expected for purely baryonic systems if these retain their stellar
remnants and hot gas.  For example, the mass of an E~galaxy may be
comprised of only 30~per cent or less of stars, the rest
  consisting of  stellar remnants and gas that cannot cool to form
new stars \citep{PB08,DKB09}, meaning that $\Upsilon=5$ would be an
underestimate in that case.  Ultra-compact dwarf galaxies, UCDs
(sometimes also understood as extremely massive star clusters), have
high stellar $M/L$ values perhaps due to a bottom-heavy IMF
\citep{MK08} or a top-heavy IMF \citep{DKB09}.

\begin{figure}
\vspace{3mm} \includegraphics[angle=0,scale=0.48]{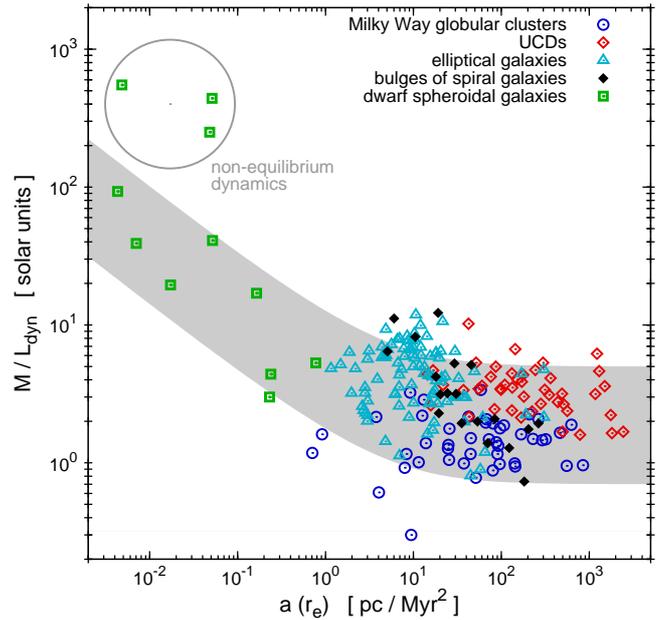}
\vspace{2mm}
\caption{The correlation between the acceleration $a(r_e)$ and the
  dynamical mass-luminosity ratio $(M/L)_{\rm dyn}$ derived assuming
  Newtonian dynamics is shown for the same objects as in
  Fig.~\ref{fig:accel}. The shaded region indicates the range in
  $(M/L)_{\rm dyn}$ as it follows directly from MOND models (without
  any parameter adjustments) using Eq.~\ref{eq:MOND}.  The graph shows
  the consistency of the data in a MONDian universe for an observer
  who interprets observations with Newtonian dynamics.  Encircled
  dwarf spheroidals outside this range (UMa, Dra, and UMi) may indicate
  non-equilibrium dynamics, either because the whole system is
  unbound, or because of unbound interloper stars among the member
  stars (see Sect.~\ref{ssec:nonnewton}). That virtually all
  pressure-supported stellar systems fall in the shaded MOND region
  suggests a successful consistency check. That is, stellar dynamics
  is MONDian rather than Newtonian on galactic scales.
\label{fig:a_correl}}
\end{figure}

By comparing the two panels in Fig.~\ref{fig:accel}, it is indeed
evident that only those systems with $a<a_o$ show non-baryonic
$(M/L)_{\rm dyn}$ values.  This is more clearly shown in
Fig.~\ref{fig:a_correl} where the MOND prediction for the range of
dynamical mass-to-light ratios measured by a Newtonist living in a
MONDian universe is plotted as a function of Newtonian
acceleration. For this figure, the MOND expectation for the
mass-to-light ratio, which an observer who thinks to live in a
Newtonian world would deduce, was calculated as follows. Adopting a
conservative value of the baryonic mass-to-light ratio $\Upsilon_{\rm
  bar}$ between 0.7 (for a globular cluster with an old metal-poor
population depleted in low-mass stars) and 5 (for an old metal-rich
population), the prediction of MOND inside the effective radius is
\citep{FB05,Angusetal09}
\begin{equation}
(M/L)_{\rm dyn \, mond} = 0.5 \times \Upsilon_{\rm bar} 
\times \left( 1+\sqrt{1+4a_o/a} \right) \ \ .
\label{eq:MOND}
\end{equation}
We note that, writing customarily $x=g/a_o$, where $g$ is the actual
full acceleration experienced by a ballistic particle (in
MOND)\footnote{In the notation applied here, the MOND formula becomes
  $a=\mu(x)\,g$, where the Newtonian acceleration $a$ is given by
  Eq.~\ref{eq:a}.}, Eq.~\ref{eq:MOND} follows from the form of the
transition MOND function \citep{Milgrom83}
\begin{equation}
\mu(x)=x/(1+x),
\label{eq:mu}
\end{equation}
which is valid up to $x\approx10$.  The theoretical transition derived
by \cite{Milgrom99} and mentioned in the Appendix would yield
virtually the same result.

The three classical dwarfs that lie outside the predicted MOND range
for $(M/L)_{\rm dyn}$ in Fig.~\ref{fig:a_correl} are UMa, Draco, and
UMi. UMa may have an anisotropic velocity dispersion \citep{Angus08};
Draco is known to be a long-standing problem for MOND, but the technique
of interloper removal developed by \cite{Serra09} could probably solve
the problem, although this particular case remains open to debate; UMi
is a typical example of a possibly out-of-equilibrium system, as it is
elongated with substructure and shows evidence of tidal tails
(D. Martinez-Delgado, priv. communication).  Ultra-faint dwarf
spheroidals are expected to be increasingly affected by this kind of
non-equilibrium dynamics, as shown to be true even for Newtonian
weak-field dynamics (\citealt{Kroupa97}, Sect.~\ref{ssec:tdgsevol}),
and even more strongly so in MOND \citep{McGaughWolf10}.

{\sl Summarising Subsect.~\ref{sec:gravdyn}}, well-developed
non-Newtonian weak-field approaches exist and have been shown to
account for galaxy properties much more succesfully than the CCM,
which would need to be extended by a dark force to account for the
observed strong coupling between DM and baryons.  All known
pressure-supported stellar systems ranging from elliptical to dwarf
satellite galaxies behave dynamically as expected in a MONDian
universe.  In DM cosmology, the association of highly non-stellar
$(M/L)_{\rm dyn}$ values with $a<a_o$ would be coincidental as it is
not built into the theory.  It is, however, natural in a MONDian
universe for observers who interpret weak-field observations with
Newtonian dynamics.

\section{Conclusions and perspectives}
\label{sec:concs}

We inhabit a Universe for which physicists seek mathematical
formulations. A successful formulation of gravitational physics, the
General Theory of Relativity (GR), requires the existence of
non-baryonic dark matter (DM) in order to account for the observed
rotation curves of galaxies and other dynamical effects in this
theory, which has Newtonian dynamics as its weak-field limit. On the
other hand, non-Newtonian weak-field gravitational theories have also
been formulated to account for the ``DM-effects'' observed in
galaxies.

Finding a definitive test that distinguishes between these two
different solutions to the problem of galactic dynamics and
cosmological structure formation is difficult. Both DM and modified
gravity are designed to solve similar problems, so the test must rely
on subtle differences between the models and the observational data.
Thus far, GR$+$DM$+\Lambda$+inflation (the CCM) accounts for the
emergence of structure on large scales, and \cite{Reyesetal2010} were
able to exclude certain versions of alternative gravitational theories
that had already been known by the respective community to be unstable
\citep{Contaldietal08}. But, as shown here, the CCM appears to have
insurmountable problems on galaxy scales such that other alternative
approaches need to be studied.  A speculative ansatz to perhaps solve
the observed near-exact DM--baryon coupling in galaxies within a
DM-Newtonian cosmology would be to extend the CCM by postulating the
existence of a {\sl dark force} (DF) leading to a
GR$+$DM$+$DF$+$$\Lambda$$+$inflation cosmology that should perhaps be
investigated in more detail in the future.  The greatest differences
between the two competing approaches (CCM versus non-Newtonian
dynamics in the weak-field limit) are expected in the weak
gravitational regime where the subtleties of non-Newtonian weak-field
dynamics are most pronounced, which is why the constituents of the
outer edges of galaxies allow the most stringent tests.

This contribution has statistically assessed whether the observed
properties of satellite galaxies in the Local Group, which are the
result of structure formation in the weak-field limit, are consistent
with the CCM.  Given that a substantial number of independent research
groups working in the traditional CDM and WDM approaches have by now
made firm statements about the dwarf satellite galaxies of the MW and
Andromeda such that the missing satellite problem is deemd to be
solved, the CCM can be further tested sensitively on these scales
within the Local Group.

Five new problems for the CCM on the scale of the Local Group and
dwarf galaxies have been uncovered: (i) the observed absence of a
mass-luminosity relation (Sect.~\ref{sec:ML}, the {\sl
  DM-mass--luminosity problem}); (ii) the mass function of luminous
galactic satellites (Sect.~\ref{sec:mfn}, the {\sl mass function of
  luminous satellite problem}); (iii) the observed relation between
the bulge mass and the number of satellites (Sect.~\ref{sec:origin},
the {\sl bulge-satellite correlation problem}); (iv) the accordance
with the Milky Way's disc-of-satellites of the recently detected
ultra-faint dwarfs (Sect.~\ref{sec:DoS}, the {\sl phase-space
  correlation problem}); and (v) the low probability that two
neighbouring MW-type DM halos contain similar MW-type disk galaxies
(Sect.~\ref{ssec:baryonic_invariant}, the {\sl
  invariant-baryonic-galaxy problem}).

It is found that the CCM is consistent with the Local Group data with
a combined probability\footnote{Summarising the likelihoods, $p$, that
  the CCM accounts for the observed data in the Local Group are in the
  individual tests: (1)~mass--luminosity data: $p_1<0.3$~per cent
  (Sec.~\ref{sec:ML}); (2)~mass function of luminous sub-halos:
  $p_2<4.5$~per cent (Sect.~\ref{sec:mfn}); (3)~bulge--satellite
  number: $p_3\approx4.4$~per cent (Sect.~\ref{sec:origin});
  (4)~a~MW-type galaxy with at least 11~satellites in a DoS:
  $p_{4}=0.4$~per cent; (5)~a~M31-type galaxy with at least 11
  satellites: $p_{5}=1.4$~per cent (Sect.~\ref{ssec:DoSDisc}). Thus,
  the combined probability that the general CCM framework accounts for
  the Local Group is $p \ll 3\times 10^{-3}$.}
%
%ML: 0.3 percent likely 
%
%MFn: 4.5 percent likely 
%
%Bulge-N: 0.64 per cent likely 
%
%Local Group: MW with 11 sats in DoS: 0.4 per cent likeley, M31 with at
%least 11 sats: 1.4 percent likely, t get Local Group with MW and M31
%is 5.6E-3 per cent likely
%
%thus, combined likelihood that the general CCM framework accounts for
%the Local Group is 4.8E-11 (e.8E-9 percent)
$p\ll 3 \times 10^{-3}$. The five problems thus appear to rather
strongly challenge the notion that the CCM successfully accounts for
galactic structure in conflict with a vast volume of reported research
(compare with \citealt{Fanelli10}). All these challenges constitute a
strong motivation for numerous future observational and theoretical
investigations. For instance, the disk of satellites will have to be
confirmed by surveys such as Pan-Starrs \citep{Burgett09} and
the Stromlo Milky Way Satellite Survey (SMS) \citep{Jerjen10}. Given
the existence of the DoS and by symmetry, the Southern hemisphere
ought to also contain about 16 satellites, such that the SMS survey is
expected to discover about 8 new southern satellites
(Fig.~\ref{fig:discfit}).  It will also be essential to refine
the correlation between bulge-mass and satellite-number with
extragalactic surveys. On the theoretical side, more inclusive
modelling is needed to address these challenges within the
CCM while, at the same time, existing viable alternatives should
be explored with more emphasis.

With this contribution, the following clues have emerged suggesting
the need for a new scenario for the origin and nature of dSph
satellite galaxies.  The observed correlation between bulge mass and
number of satellites suggests that a link between these two quantities
may exist. The phase-space correlation of the classical and
ultra-faint satellite galaxies implies that angular momentum
conservation played an important role in establishing the satellite
distribution. Given that bulges form in dissipational encounters,
during which angular-momentum conservation rearranges matter on
Galactic scales to be in highly correlated phase-space structures
(tidal arms), a natural path thus opens to understand the likely
origin of satellite galaxies.  Already in the 1970's a tidal origin
for dwarf spheroidal galaxies was suggested, based on their
arrangement around the Milky Way (Sect.~\ref{sec:tdgs}).  This
solution does imply, however, that the dSph galaxies are ancient TDGs
and not DM sub-haloes.  Furthermore, by logical implication, dE
galaxies would also be TDGs (Sec.~\ref{ssec:dE}). This would imply
that the vast majority of $\simless 10^{10}\,M_\odot$ DM sub-halos are
unable to make stars. This, however, would be in conflict with all the
CCM computations (the Fritz Zwicky Paradox) available to date {\sl to
  the extend that the CCM would have to be discarded in favour of a
  universe without cold or warm DM}.  In this case, the non-Keplerian
rotation curves of galaxies and other DM effects additionally suggest
that we live in a non-Newtonian weak-field framework within which
galaxies would be pure baryonic objects\footnote{Given that Newton
  derived the gravitational $1/r^2$ law over a very limited physical
  range (Solar System), while with the Local Group gravitational
  physics is probed on a length scale nearly eight orders of magnitude
  larger and in a much weaker field regime, it need not be surprising
  that an adjusted gravitational law is needed.}.

This scenario would naturally solve problems (iii) and (iv), while it
would not imply a ``dynamical mass"-luminosity relation if the dwarfs
are out of equilibrium, so could possibly solve problem (i). For
purely baryonic galaxies, problem (ii) would not exist anymore by
definition. Problem~(v) would also vanish naturally.  What is more,
while in the CCM the association of highly non-stellar $(M/L)_{\rm
  dyn}$ values with $a<a_o$ would be coincidental because it is not built
into the theory, it is natural in a non-Newtonian universe for
weak-field observers who interpret observations with Newtonian
dynamics. Noteworthy is that the same statement can be made for the
Tully-Fisher scaling relation for rotationally-supported galaxies
\citep{TF77,McGaugh05b,CO09} as well as the newly found scaling
relation of \cite{Gentileetal09} and \cite{Milgrom09}. The supposed
mass-deficit seen in young rotating and gaseous TDGs (such as those of
NGC~5291) constitutes independent empirical evidence towards this same
statement. Young tidal dwarf galaxies (TDG), which should be devoid of
collisionless DM, appear to nevertheless exhibit a mass-discrepancy in
Newtonian dynamics. This is a significant problem for the DM
hypothesis, but it is naturally explained by MOND \citep{Gentile07,
  Milgrom07}. Also, while the high Bullet-cluster velocity is hard to
account for in the CCM, it is natural in MOND (Sect.~\ref{sec:introd},
\ref{sec:gravdyn} and~\ref{ssec:nonNewt}).
And, it has already been noted by \cite{Sanders99} that the
    dynamical-mass -- baryon-mass discrepancy observed in galaxy
    clusters is nearly removed in MONDian dynamics.

{\sl It would thus appear that within the non-Newtonian weak-field
  framework a much more complete, self-consistent, and indeed simpler
  understanding of the Galaxy's satellites as well as of major
  galaxies may be attained, than within the CCM.}

However, to affirm this statement, this alternative cosmological
scenario will have to be investigated in as much detail as is now
available for the CCM in order to perform equivalent tests as
presented here for the DM hypothesis and to ascertain which of the
non-Newtonian weak-field dynamics theories (and which versions of the
theories) can most successfully account for the physical world. Models of
merging gas-rich disc galaxies need to be computed in MOND, for example, to
study how the formation of TDGs proceeds and how the number of
satellites thus formed correlates with the bulge that forms as a
result of the encounter. These populations of satellites associated
with globular clusters that formed along with them would naturally
appear in (more than one) closely related planes explaining the
\cite{LL95} streams, because a gas-rich galaxy pair undergoes many
close encounters in MOND, each spawning some TDGs and globular
clusters, before perhaps finally merging.

\begin{figure}
\vspace{3mm}
\includegraphics[angle=0,scale=0.5]{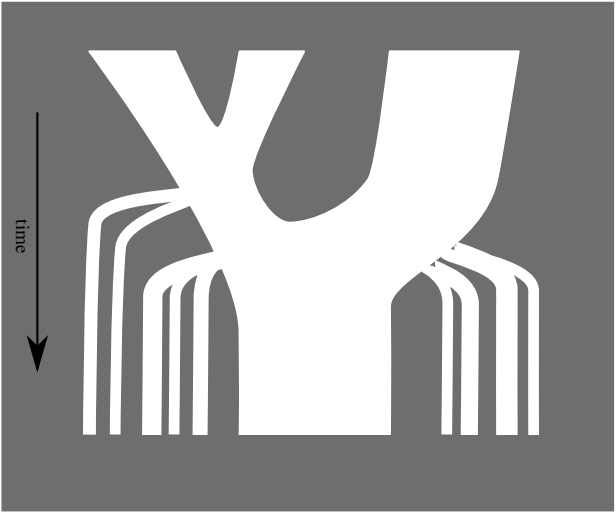}
\vspace{2mm}
\caption{A new cosmological structure formation framework: the
  mangrove merger tree. In a modified-Newtonian weak-field framework,
  purely baryonic galaxies merge thereby spawning new dwarf galaxies
  giving rise to the morphology-density relation (adapted from
  \citealt{Metz08a}).
\label{fig:mangrove}}
\end{figure}

Figure~\ref{fig:mangrove} schematically depicts the structure
formation scenario in this non-Newtonian weak-field framework: while
purely baryonic galaxies would merge, these events would spawn dwarf
galaxies such that a density--morphology relation would be established
(more dE galaxies in denser environments, \citealt{OT00}).

The MONDian modelling by \cite{TC08} and \cite{CT09} has already shown
that TDGs are produced during gas-dissipational galaxy mergers, and
that the interaction times between galaxies are much longer, while the
number of mergers is smaller than in a DM
universe. Hence, the number of observed galaxy encounters would 
be given foremost by the long time scale of merging 
and thus by more close galaxy-galaxy encounters per merging event
rather than on a high number of mergers.

This would imply that compact galaxy groups do not evolve statistically
over more than a crossing time. In contrast, assuming DM-Newtonian
dynamics to hold, the merging time scale would be about one crossing
time because of dynamical friction in the DM halos such that compact
galaxy groups ought to undergo significant merging over a crossing
time. The lack of significant evolution of compact groups, if verified
observationally, would appear not to be explainable if DM dominates
galaxy dynamics. Analyses of well-studied compact groups indeed indicate
this to be the case \citep{Presottoetal10}.

Thus, many observational problems may be solved uncontrived by
adopting non-Newtonian weak-field dynamics, and perhaps this was, in
the end, the most self evident explanation to the discovery of
non-Keplerian rotation curves by \cite{RF70}\footnote{On 19~June~2009,
  the final day of the conference "Unveiling the Mass: Extracting and
  Interpreting Galaxy Masses" in Kingston, Ontario, in honour of the
  career of Vera Rubin, PK asked her whether she would be very
  dismayed if her discovery that galaxies have non-Keplerian rotation
  curves would not be due to dark mater but rather non-Newtonian
  weak-field dynamics. Prof.~Rubin replied that she would in fact be
  delighted, since the non-Keplerian rotation curves are an empirical
  observation of hitherto not understood physics, and one needs to
  keep an open mind in seeking solutions.}.

\acknowledgements{This work was suported by the Alexander von Humboldt
  Foundation (BF), and by the German Research Foundation (DFG) through
  grants KR1635/18-1 and HE1487/36-2 within the priority programme
  1177 ``Witnesses of Cosmic History: Formation and Evolution of Black
  Holes, Galaxies and Their Environment'', and a grant from the
  DAAD-Go8 Germany Australia Joint Research co-operative scheme.  We
  acknowledge useful discussions with Iskren Georgiev, Anton
  Ippendorf, Antonino Del Popolo and Beth Willman. We thank Jelte de
  Jong for allowing us to use the image from \cite{Coleman2007} in our
  Fig.~\ref{fig:hercules}.}

\bibliographystyle{aa}

\newpage

\noindent
{\large\bf Appendix: A brief review of MOND and MOG and Milgrom's
  proposition on the possible physical origin and value of
  $a_0$}
\label{app:a0}

\vspace{3mm}

\noindent
Theoretical approaches trying to embed MOND within a Lorentz-covariant
framework \citep{Bekenstein04,Sanders05, Zlosnik07,Zhao08,
  Brun08,Blanchet09, EF09,Skordis09, Milgrom09b} are currently under
intense scrutiny, and a quasi-linear formulation of MOND has been
discovered only recently \citep{Milgrom10, ZB10}, which appears to
allow easier access to N-body calculations.  

However, none of these theories is (yet) fully satisfactory from a
fundamental point of view (see e.g. \citealt{Contaldietal08, Brun08,
  Reyesetal2010}) and moreover none of them explains (yet) why the
acceleration threshold, $a_0$, which is the single parameter of MOND
(adjusted by fitting to one single system), is about
$c\sqrt{\Lambda/3}$ (where $\Lambda$ is the cosmological constant and
$c$ the speed of light), or that $a_0\approx c H_0/2\,\pi$, where
$H_0$ is the current Hubble constant. They also require a transition
function, $\mu(x)$ (e.g. Eq.~\ref{eq:mu}), from the Newtonian to the
modified regime, a function not (yet) rooted in the theory.

A possible explanation of the coincidence $a_0\approx
c\sqrt{\Lambda/3}$ and a theoretically-based transition function are
suggested by \cite{Milgrom99}. In Minkowski (flat) space-time, an
accelerated observer sees the vacuum as a thermal bath with a
temperature proportional to the observer's acceleration
\citep{Unruh}. This means that the inertial force in Newton's second
law can be defined to be proportional to the Unruh temperature.  On
the other hand, an accelerated observer in a de~Sitter universe
(curved with a positive cosmological constant $\Lambda$) sees a
non-linear combination of the \cite{Unruh} vacuum radiation and of the
\cite{GibbonsHawking} radiation due to the cosmological horizon in the
presence of a positive $\Lambda$.  \cite{Milgrom99} then defines
inertia as a force driving such an observer back to equilibrium as
regards the vacuum radiation (i.e. experiencing only the
Gibbons-Hawking radiation seen by a non-accelerated observer).
Observers experiencing a very small acceleration would thus see an
Unruh radiation with a low temperature close to the Gibbons-Hawking
one, meaning that the inertial resistance defined by the difference
between the two radiation temperatures would be smaller than in
Newtonian dynamics, and thus the corresponding acceleration would be
larger. This is given precisely by the MOND formula of
\cite{Milgrom83} with a well-defined transition-function $\mu(x)$, and
$a_o=c\,(\Lambda/3)^{1/2}$. Unfortunately, no covariant version (if at
all possible) of this approach has been developed yet.

The theoretical basis of the MOG approach relies on chosen values of
integration constants in solving the equations of the theory.  This
approach seems to work well from an observational point of view, but
it's fundamental basis needs further research, as is also the case for
MOND. 
It is noteworthy that a formulation of MOG in terms of
scalar, vector, and tensor fields \citep{Moffat06} may possibly hint at a
convergence with the \cite{Bekenstein04} tensor-vector-scalar
theory of gravity.

\end{document}